\documentclass[iop, numberedappendix]{emulateapj} 

\usepackage{graphicx}
\usepackage{pslatex}
\usepackage{amsmath, amsthm, amssymb}
\usepackage{natbib}
\usepackage{epsfig}
\usepackage{subfigure}

\shorttitle{Optimized Multi-Frequency Spectra}
\shortauthors{Mirocha et al.}

\begin{document}
    
\newcommand{\HI}{\text{H} {\textsc{i}}}
\newcommand{\HII}{\text{H} {\textsc{ii}}}
\newcommand{\HeI}{\text{He} {\textsc{i}}}
\newcommand{\HeII}{\text{He} {\textsc{ii}}}
\newcommand{\HeIII}{\text{He} {\textsc{iii}}}

\newcommand{\nH}{n_{\text{H}}}
\newcommand{\nHe}{n_{\text{He}}}
\newcommand{\nHI}{n_{\text{H } \textsc{i}}}
\newcommand{\nHII}{n_{\text{H } \textsc{ii}}}
\newcommand{\nHeI}{n_{\text{He } \textsc{i}}}
\newcommand{\nHeII}{n_{\text{He } \textsc{ii}}}
\newcommand{\nHeIII}{n_{\text{He } \textsc{iii}}}
\newcommand{\nel}{n_{\text{e}}}  
\newcommand{\ntot}{n_{\text{tot}}}

\newcommand{\NHI}{N_{\text{H } \textsc{i}}}
\newcommand{\NHeI}{N_{\text{He } \textsc{i}}}
\newcommand{\NHeII}{N_{\text{He } \textsc{ii}}}

\newcommand{\xHI}{x_{\text{H } \textsc{i}}}
\newcommand{\xHII}{x_{\text{H } \textsc{ii}}}
\newcommand{\xHeI}{x_{\text{He } \textsc{i}}}
\newcommand{\xHeII}{x_{\text{He } \textsc{ii}}}
\newcommand{\xHeIII}{x_{\text{He } \textsc{iii}}}

\newcommand{\ionHI}{\Gamma_{\text{H } \textsc{i}}}
\newcommand{\ionHeI}{\Gamma_{\text{He } \textsc{i}}}
\newcommand{\ionHeII}{\Gamma_{\text{He } \textsc{ii}}}
\newcommand{\ionsecHI}{\gamma_{\text{H } \textsc{i}}}
\newcommand{\ionsecHeI}{\gamma_{\text{He } \textsc{i}}}
\newcommand{\ionsecHeII}{\gamma_{\text{He } \textsc{ii}}}
\newcommand{\ioncollHI}{\beta_{\text{H } \textsc{i}}}
\newcommand{\ioncollHeI}{\beta_{\text{He } \textsc{i}}}
\newcommand{\ioncollHeII}{\beta_{\text{He } \textsc{ii}}}
\newcommand{\recHII}{\alpha_{\text{H } \textsc{ii}}}
\newcommand{\recHeII}{\alpha_{\text{He } \textsc{ii}}}
\newcommand{\recHeIII}{\alpha_{\text{He } \textsc{iii}}}

\newcommand{\heatHI}{\mathcal{H}_{\text{H } \textsc{i}}}
\newcommand{\heatHeI}{\mathcal{H}_{\text{He } \textsc{i}}}
\newcommand{\heatHeII}{\mathcal{H}_{\text{He } \textsc{ii}}}

\newcommand{\cooldielHeII}{\omega_{\text{He } \textsc{ii}}}

\newcommand{\PhiHI}{\Phi_{\text{H } \textsc{i}}}
\newcommand{\PhiHeI}{\Phi_{\text{He } \textsc{i}}}
\newcommand{\PhiHeII}{\Phi_{\text{He } \textsc{ii}}}
\newcommand{\PsiHI}{\Psi_{\text{H } \textsc{i}}}
\newcommand{\PsiHeI}{\Psi_{\text{He } \textsc{i}}}
\newcommand{\PsiHeII}{\Psi_{\text{He } \textsc{ii}}}

\newcommand{\nnu}{$n_{\nu}$}
\newcommand{\ncol}{N_i}
\newcommand{\gHI}{\Gamma_{\text{HI}}}  
\newcommand{\gHeI}{\Gamma_{\text{HeI}}}
\newcommand{\gHeII}{\Gamma_{\text{HeII}}}
\newcommand{\aHII}{\alpha_{\text{HII}}}  
\newcommand{\aHeII}{\alpha_{\text{HeII}}}  
\newcommand{\aHeIII}{\alpha_{\text{HeIII}}}
\newcommand{\bHI}{\beta_{\text{HI}}} 
\newcommand{\bHeI}{\beta_{\text{HeI}}}  
\newcommand{\bHeII}{\beta_{\text{HeII}}}  
\newcommand{\xiHeII}{\xi_{\text{HeII}}}
\newcommand{\kB}{k_{\text{B}}}
\newcommand{\fheat}{f^{\text{heat}}}
\newcommand{\fion}{f_i^{\text{ion}}}
\newcommand{\Lbol}{\mathcal{L}_{\text{bol}}}
\newcommand{\spec}{\mathcal{N}}
\newcommand{\Heat}{\mathcal{H}}
\newcommand{\trec}{$t_{\text{rec}}$}
\newcommand{\Lbox}{L_{\mathrm{box}}}
\newcommand{\dx}{\Delta x}
\newcommand{\dd}{\text{d}}
\newcommand{\Htwo}{\mathrm{H}_2}
\newcommand{\drIF}{$\Delta r_{\mathrm{IF}}$}
\newcommand{\dTb}{$\delta T_b$}
\newcommand{\Nvec}{\mathbf{N}}
\newcommand{\sh}{\mathrm{sh}}

\title {Optimized Multi-Frequency Spectra for Applications in Radiative Feedback and Cosmological Reionization}
\author{Jordan Mirocha$^{\dagger}$, Stephen Skory, Jack O. Burns}
\affil{Center for Astrophysics and Space Astronomy, University of Colorado, Campus Box 389, Boulder, CO 80309}
\affil{The NASA Lunar Science Institute, NASA Ames Research Center, Moffett Field, CA 94035, USA}
\email{$^{\dagger}$jordan.mirocha@colorado.edu}
\and
\author{John H. Wise}
\affil{Center for Relativistic Astrophysics, School of Physics, Georgia Institute of Technology, Atlanta, GA
30332}

\begin{abstract}
The recent implementation of radiative transfer algorithms in numerous
hydrodynamics codes has led to a dramatic improvement in studies of feedback
in various astrophysical environments. However, because of methodological
limitations and computational expense, the spectra of radiation sources are
generally sampled at only a few evenly-spaced discrete emission frequencies.
Using one-dimensional radiative transfer calculations, we investigate the
discrepancies in gas properties surrounding model stars and accreting black
holes that arise solely due to spectral discretization. We find that even in
the idealized case of a static and uniform density field, commonly used
discretization schemes induce errors in the neutral fraction and temperature
by factors of two to three on average, and by over an order of magnitude in
certain column density regimes. The consequences are most severe for radiative
feedback operating on large scales, dense clumps of gas, and media consisting
of multiple chemical species. We have developed a method for optimally
constructing discrete spectra, and show that for two test cases of interest,
carefully chosen four-bin spectra can eliminate errors associated with
frequency resolution to high precision. Applying these findings to a fully
three-dimensional radiation-hydrodynamic simulation of the early universe, we
find that the \HII\ region around a primordial star is substantially altered in
both size and morphology, corroborating the one-dimensional prediction that
discrete spectral energy distributions can lead to sizable inaccuracies in the
physical properties of a medium, and as a result, the subsequent evolution and
observable signatures of objects embedded within it.
\end{abstract} 
\subjectheadings {dark ages, reionization, first stars --- methods: numerical --- radiative transfer}

\section{INTRODUCTION} \label{sec:Introduction}
Energy injection by radiative processes fundamentally changes the evolution of
astrophysical systems, whether it be in the context of star formation, galaxy
evolution, or the growth of super--massive black holes (SMBHs). For instance,
ultraviolet photons from the universe's first stars \citep[Population III
(PopIII) stars;][]{Abel2002b} photo-dissociate the primary coolant ($\Htwo$)
that first enabled their formation. Very recent radiation-hydrodynamic
calculations of PopIII stars find that PopIII star masses may be limited by
proto-stellar radiative feedback, perhaps explaining the lack of evidence for
exotic pair instability supernovae in the early universe \citep{Hosokawa2011}.
Conventional metal line cooling driven star formation can be affected by
radiative feedback as well. \citet{Krumholz2006} showed that photo-heating
around newly formed stars can strongly suppress fragmentation in surrounding
proto-stellar clouds, while \citet{Dale2005} see both positive and negative
feedback operating in radiation-hydrodynamic simulations of star cluster
formation. Radiative feedback could also be a barrier to efficient black hole
(BH) growth in the early universe \citep{Alvarez2009}, as X-rays from
accreting BHs efficiently photo-heat surrounding gas, leading to smaller
Bondi--Hoyle accretion rates \citep{Bondi1944}.

The mere presence of ionizing/dissociating photons ensures a change in the
chemical and thermal state of a gas, though the magnitude of these changes
hinges squarely on the number of photons propagating through the gas and their
spectral energy distribution (SED). Holding the bolometric luminosity of a
radiation source constant, even subtle changes in the SED can lead to
noticeable differences in the properties of the surrounding medium. For
example, adjusting the X-ray power-law index of a BH accretion spectrum
results in ionization fronts which differ by factors of $\approx$ 2-3 in
radius, and temperature profiles varying by $10^2$-$10^3$K on scales of
several hundred kpc \citep{Thomas2008}. Simply truncating the emission of
identical X-ray SEDs at harder energies (0.4 keV rather than 0.2 keV) causes a
drastic reduction in heating, ionized fractions, and $\Htwo$ fractions
surrounding `miniquasars' at high redshift \citep{Kuhlen2005}.

Unfortunately, not all radiative transfer algorithms are able to represent
radiation sources with continuous SEDs, or perhaps cannot afford the
additional computational expense associated with the frequency dependence of
the radiative transfer equation. The natural first step is to represent
sources as monochromatic emitters, choosing an emission frequency
characteristic of the full SED. Some authors have improved upon the
monochromatic treatment using `multi-group' methods, which average SED
properties and absorption cross-sections over one or more frequency bandpasses
\citep{Gnedin2001,Aubert2008}, while others have sampled continuous SEDs at
\nnu\ frequencies, which are generally evenly spaced bins (in linear or
log-space) between the hydrogen ionization threshold and an upper frequency
cutoff. In either case, there is no clear method of deciding how many
frequency-averaged bandpasses or discrete emission frequencies are required
for a given problem, and though the standard multi-group treatment is
physically motivated, it does not guarantee that the photo-ionization and
photo-heating rates are adequately reproduced as a function of column density.

Frequency resolution has recently been studied in radiation-hydrodynamic
settings by \citet{Wise2011} and \citet{Whalen2008}. \citet{Wise2011} find
that for the expansion of an \HII\ region around a $10^5$ K blackbody source
in a hydrogen-only medium, the density, temperature, velocity, and ionization
profiles are well converged for $n_{\nu} \ge 4$. Use of a monochromatic
spectrum for this problem introduces significant errors since all photons are
absorbed at a characteristic column density, whereas multi-frequency
treatments achieve some column density dependent behavior and can thus mimic
the behavior of a truly continuous spectrum. \citet{Whalen2008} studied the
effects of frequency resolution in the setting of I-front instabilities, and
did not achieve convergence until $n_{\nu} \ge 80$ (logarithmically spaced
between 13.6 and 90 eV).
 
The convergence for the test of \citet{Wise2011} using only four frequency
bins is reassuring, though the prospects for convergence are less clear if one
were interested in the absorption processes of multiple chemical species,
ionization and heating due to X-rays and their energetic secondary
photo-electrons \citep{Shull1985, Furlanetto2010}, or inhomogeneous media.
\citet[][hereafter KH08]{Kramer2008} briefly compared monochromatic and
continuous treatments of absorbed power-law X-ray sources in a study of
ionization front thickness around high-$z$ quasars (the I-front thickness is a
potentially powerful indirect probe of the ionizing spectrum of high-$z$
quasars). The hydrogen and helium I-front thickness is expected to grow over
the lifetime of a quasar given the discrepancy in evolution timescales between
the largest and smallest scales. At small radii, photo-ionization equilibrium
is reached quickly since ionizing photons are abundant, whereas geometrical
dilution and attenuation of the initial radiation field slow ionization
evolution considerably on large scales, effectively `stretching out' the
I-fronts of hydrogen and helium with time. A monochromatic representation of
the quasar SED leads to a reduction in this effect, but also leads to severe
errors in the overall ionization structure (see Figure 3 of KH08). These
errors are of the same order of magnitude as those resulting from the neglect
of physical effects, such as ionization via helium recombination photons
(KH08, Figure 6), or ionization from secondary electrons (KH08, Figure 7).
These effects are likely important in studies of radiative feedback from stars
and active galactic nuclei (AGNs), and most certainly in efforts to simulate
cosmological reionization. An effort must be made to ensure that the SEDs used
in numerical simulations accurately reflect the properties of their continuous
analogs, especially if it is spectrum-dependent effects in which we are most
interested.

We will focus on the following questions in this paper. How significant are
the errors in the temperature and ionization state of a medium that arise
solely due to the discretization of SEDs? How many frequencies are required to
minimize such errors, where must they be positioned in frequency-space, and
how should their relative luminosities be apportioned? For what numerical
methods is it possible to represent sources with continuous SEDs, or are there
perhaps advantages in discretizing SEDs, even when it is not required by the
algorithm of choice? Answers to these questions may lead to revised
interpretations of previous studies which used discrete radiation fields, but
more importantly, will reduce the guesswork involved in discretizing SEDs, and
promote frequency resolution to the same status as spatial, temporal, and mass
resolution, which are more easily selected on a problem-by-problem basis.

In Section \ref{sec:Framework} we will introduce the one-dimensional radiative
transfer framework used to obtain the solutions presented in later sections.
In Section \ref{sec:Consequences}, we quantitatively assess the accuracy with
which multi-frequency calculations reproduce the ionization and heating
profiles of continuous SEDs. Section \ref{sec:Methods} is devoted to
introducing a technique for optimally selecting discrete SED templates, and
Section \ref{sec:Results} will present the results obtained with this method,
including applications to one-dimensional and fully three-dimensional
radiation-hydrodynamic calculations. Discussion and conclusions can be found
in Sections \ref{sec:Discussion} and \ref{sec:Conclusions}, respectively.
Validation of the radiative transfer code used for this work and further
details regarding the optimization algorithm can be found in the Appendix.

\section{RADIATIVE TRANSFER FRAMEWORK} \label{sec:Framework}
One dimensional radiative transfer calculations around point sources have been
used to model cosmological reionization \citep{Fukugita1994}, the thickness of
quasar ionization fronts (KH08), the time-evolution of ionization and heating
around first stars, galaxies, and quasars \citep{Thomas2008,Venkatesan2011},
and their associated observable signatures. Given that our focus is on
frequency resolution, it would be unnecessary to perform calculations in a
more complex setting than this, with additional unrelated physics. As a
result, our one-dimensional methods strongly resemble those used by previous
authors, though for completeness, we will reiterate the aspects of these
methods most pertinent to the problem at hand.

In general, the chemical and thermal evolution of gas surrounding a radiation
source is governed by a set of differential equations describing the number
densities of all ions and the temperature of the gas. Assuming a medium
consisting of hydrogen and helium only, we first solve for the abundances of
each ion via
\begin{align}
    \frac{d \nHII}{dt} & = (\ionHI + \ionsecHI + \ioncollHI \nel) \nHI - \recHII \nel \nHII   \label{eq:HIIRateEquation} \\ 
    \frac{d \nHeII}{dt} & = (\ionHeI + \ionsecHeI + \ioncollHeI \nel) \nHeI \nonumber + \recHeIII \nel \nHeIII \\  & - (\ioncollHeII + \recHeII + \xiHeII) \nel \nHeII \label{eq:HeIIRateEquation} \\ 
    \frac{d \nHeIII}{dt} & = (\ionHeII + \ionsecHeII + \ioncollHeII \nel) \nHeII  - \recHeIII \nel \nHeIII . \label{eq:HeIIIRateEquation}
\end{align}
Each of these equations represents the balance between ionizations of species
\HI, \HeI, and \HeII, and recombinations of \HII, \HeII, and
\HeIII. Associating the index $i$ with absorbing species, $i = $\HI, \HeI,
\HeII, and the index $i^{\prime}$ with ions, $i^{\prime} = $\HII, \HeII,
\HeIII, we define $\Gamma_i$ as the photo-ionization rate coefficient,
$\gamma_i$ as the secondary ionization rate coefficient, $\alpha_{i^{\prime}}$
($\xi_{i^{\prime}}$) as the case-B (dielectric) recombination rate
coefficients, $\beta_i$ as the collisional ionization rate coefficients, and
$\nel = \nHII + \nHeII + 2\nHeIII$ as the number density of electrons.

At each time step, we also solve for the temperature evolution, $dT_k/dt$,
which is given by
\begin{align}
    \frac{3}{2}\frac{d}{dt}\left(\frac{\kB T_k \ntot}{\mu}\right) & = \fheat  \sum_i n_i \Heat_i - \sum_i \zeta_i \nel n_i - \sum_{i^{\prime}} \eta_{i^{\prime}} \nel n_{i^{\prime}} \nonumber \\ & - \sum_i \psi_i \nel n_i - \cooldielHeII \nel \nHeII \label{eq:TemperatureEvolution} 
\end{align}
where $\Heat_i$ is the photo--electric heating rate coefficient (due to
electrons previously bound to species $i$), $\cooldielHeII$ is the dielectric
recombination cooling coefficient, and $\zeta_i$, $\eta_{i^{\prime}}$, and
$\psi_i$ are the collisional ionization, recombination, and collisional
excitation cooling coefficients, respectively. The constants in Equation
(\ref{eq:TemperatureEvolution}) are the total number density of baryons,
$\ntot = n_\mathrm{H} + n_{\mathrm{He}} + \nel$, the mean molecular weight,
$\mu$, Boltzmann's constant, $\kB$, and the fraction of secondary electron
energy deposited as heat, $\fheat$. We use the formulae in Appendix B of
\citet{Fukugita1994} to compute the values of $\alpha_i$, $\beta_i$, $\xi_i$,
$\zeta_i$, $\eta_{i^{\prime}}$, $\psi_i$, and $\cooldielHeII$.

The most critical aspect of propagating the radiation field in our
one-dimensional simulations is computing the ionization ($\Gamma_i$,
$\gamma_{i}$) and heating ($\Heat_i$) rate coefficients accurately. In order
to directly relate our results to fully three-dimensional radiative transfer
calculations, we have chosen to adopt a photon-conserving (PC) algorithm
nearly identical to those employed by several widely used codes, like
\textit{C$^2$Ray} \citep{Mellema2006,Friedrich2012}, and \textit{Enzo}
\citep{Wise2011}. Our code is able to compute $\Gamma_i$, $\gamma_{i}$, and
$\Heat_i$ in a non-photon-conserving (NPC) fashion as well, to enable
comparison with previous one-dimensional work such as \citet{Thomas2008}. The
two formalisms are equivalent in the limit of very optically thin cells, a
condition that can be met easily in one-dimensional calculations but is rarely
computationally feasible in three dimensions. For NPC methods, if the optical
depth of an individual cell is substantial, the number of ionizations in that
cell will \textit{not} equal the number of photons absorbed for that cell,
i.e., photon number will not be conserved. This problem was remedied by
\citet{Abel1999}, who inferred the number of photo-ionizations of species $i$
in a cell from the radiation incident upon it and its optical depth,
\begin{equation}
    \Delta \tau_{i,\nu} = n_i \sigma_{i,\nu} \Delta r .
\end{equation}    
It is most straightforward to imagine our one-dimensional grid as a collection
of concentric spherical shells, each having thickness $\Delta r$ and volume
$V_{\sh}(r) = 4 \pi [(r + \Delta r)^3 - r^3] / 3$, where $r$ is the distance
between the origin and the inner interface of each shell. The ionization and
heating rates can then be related to the number of absorptions in any given
shell (thus preserving photon number), as
\begin{align}
    \Gamma_i & = A_i \int_{\nu_i}^{\infty} I_{\nu} e^{-\tau_{\nu}} \left(1 - e^{-\Delta \tau_{i,\nu}}\right) \frac{d\nu}{h\nu} \label{eq:PhotoIonizationRate} \\
    \gamma_{ij} & = A_j \int_{\nu_j}^{\infty} \left(\frac{\nu - \nu_j}{\nu_i}\right) I_{\nu} e^{-\tau_{\nu}} \left(1 - e^{-\Delta \tau_{j,\nu}}\right) \frac{d\nu}{h\nu} \label{eq:SecondaryIonizationRate} \\
    \Heat_i & = A_i \int_{\nu_i}^{\infty} (\nu - \nu_i) I_{\nu} e^{-\tau_{\nu}} \left(1 - e^{-\Delta \tau_{i,\nu}}\right) \frac{d\nu}{\nu} , \label{eq:HeatingRate}
\end{align}    
where we have defined the normalization constant $A_i \equiv
L_{\mathrm{bol}}/n_i V_{\sh}(r)$, and denote the ionization threshold energy
for species $i$ as $h\nu_i$. $I_{\nu}$ represents the SED of radiation
sources, and satisfies $\int_{\nu} I_{\nu} d\nu = 1$, such that
$L_{\mathrm{bol}} I_{\nu} = L_{\nu}$.

Equation (\ref{eq:SecondaryIonizationRate}) represents ionizations of species
$i$ due to fast secondary electrons from photoionizations of species $j$,
which has number density $n_j$, and ionization threshold energy, $h\nu_j$.
$\fion$ is the fraction of photo-electron energy deposited as ionizations of
species $i$. In the remaining sections we only include the effects of
secondary electrons when considering X-ray sources, which emit photons in the
range $10^2\mathrm{eV} < E < 10^4\mathrm{eV}$. In this regime, the values of
$\fheat$ and $\fion$ computed via the formulae of \citet{Shull1985} are
sufficiently accurate, but for radiation at lower energies where $\fheat$ and
$\fion$ have a stronger energy dependence, the fitting formulae of
\citet{Ricotti2002} or the lookup tables of \citet{Furlanetto2010} would be
more appropriate. The total secondary ionization rate for a given species,
$\gamma_i$, is the sum of ionizations due to the secondary electrons from all
species, $\gamma_i = \fion \sum_j \gamma_{ij} n_j / n_i$.

The optical depth, $\tau_{\nu} = \tau_{\nu}(r)$, in the above equations is
the total optical depth at frequency $\nu$ due to all absorbing species, i.e.,
\begin{align}
    \tau_{\nu}(r) & = \sum_i \int_0^r \sigma_{i,\nu} n_i(r^{\prime}) dr^{\prime} \nonumber \\
               & = \sum_i \sigma_{i,\nu} \ncol(r) \label{eq:OpticalDepth}
\end{align}
where $\ncol$ is the column density of species $i$ at distance $r$ from the
source. We calculate the bound--free absorption cross-sections using the fits
of \citet{Verner1996} throughout.

The values of $\Gamma_i$, $\gamma_i$, and $\Heat_i$ are completely
predetermined for a given radiation source, and as a result, can be tabulated
as a function of column density to avoid evaluating the integrals in these
expressions numerically `on-the-fly' as a simulation runs
\citep[e.g.,][]{Mellema2006,Thomas2008}. Isolating the frequency-dependent
components of Equations
(\ref{eq:PhotoIonizationRate})--(\ref{eq:HeatingRate}), we can define the
integrals
\begin{align}
    \Phi_i (\tau_{\nu}) & \equiv \int_{\nu_i}^{\infty} I_{\nu} e^{-\tau_{\nu}} \frac{d\nu}{h\nu} \label{eq:PHI} \\
    \Psi_i (\tau_{\nu}) & \equiv \int_{\nu_i}^{\infty} I_{\nu} e^{-\tau_{\nu}} d\nu \label{eq:PSI},
\end{align}
allowing us to re-express the rate coefficients as
\begin{align}
    \Gamma_i & = A_i \left[\Phi_i(\tau_{\nu}) - \Phi_i(\tau_{i,\nu}^{\prime}) \right] \label{eq:Gamma_PhiPsi} \\
    \gamma_{ij} & = \frac{A_j}{h\nu_i} \left\{\Psi_j(\tau_{\nu}) - \Psi_j(\tau_{j,\nu}^{\prime}) - h \nu_j \left[\Phi_j(\tau_{\nu}) - \Phi_j(\tau_{j,\nu}^{\prime}) \right] \right\}  \label{eq:gamma_PhiPsi} \\
    \Heat_i & = A_i \left\{\Psi_i(\tau_{\nu}) - \Psi_i(\tau_{i,\nu}^{\prime}) - h\nu_i \left[\Phi_i(\tau_{\nu}) - \Phi_i(\tau_{i,\nu}^{\prime}) \right] \right\} \label{eq:Heat_PhiPsi},  
\end{align}
where $\tau_{i,\nu}^{\prime} \equiv \tau_{\nu} + \Delta \tau_{i, \nu}$. Later
references to ``continuous SEDs'' signify use of this technique, where the
integral values $\Phi_i$ and $\Psi_i$ are computed over a column density
interval of interest a priori using a Gaussian quadrature technique,
rather than on-the-fly via discrete summation.

Tabulating Equations (\ref{eq:PHI}) and (\ref{eq:PSI}) grants a significant
speed-up computationally, but also forms the basis of our frequency resolution
optimization strategy (Section \ref{sec:Methods}). Note, however, that in
general the dimensionality of these lookup tables is equal to the number of
absorbing species (through $\Delta \tau_{i, \nu}$), so the tables for
simulations including hydrogen only are one dimensional, while those including
hydrogen and helium are three dimensional. If we chose to adopt the secondary
electron treatment of \citet{Ricotti2002} or \citet{Furlanetto2010}, our
lookup tables would inherit an additional dimension, as the secondary
ionization and heating factors $\fion$ and $\fheat$ would depend both on
photon energy and the hydrogen ionized fraction, $\xHII$.

Equations (\ref{eq:Gamma_PhiPsi})--(\ref{eq:Heat_PhiPsi}) are completely
general for PC algorithms, whether the source SEDs are discrete or continuous
--- the only difference being for discrete SEDs, the integrals in Equations
(\ref{eq:PHI}) and (\ref{eq:PSI}) become sums over the number of discrete
emission frequencies, $n_{\nu}$. In practice, computing $\Gamma_i$,
$\gamma_i$, and $\Heat_i$ is more straightforward for sources with discrete
SEDs, as we can simply count the number of ionizations caused by photons at
each individual frequency, and convert this into the amount of excess electron
kinetic energy available for further heating and ionization. When testing the
accuracy of discrete solutions in later sections we employ this method, where
radiation is emitted at $n_{\nu}$ frequencies, with each frequency $\nu_n$
carrying a fraction $I_n$ of the source's bolometric luminosity. The
photoionization and heating coefficients can then be expressed as
\begin{align}
    \Gamma_{i,n} & = \frac{A_i I_n}{h \nu_n} e^{-\tau_{\nu_n}}(1 - e^{-\Delta \tau_{i,\nu_n}}) \label{eq:Gamma_simple}\\
    \gamma_{ij,n} & = \Gamma_{j,\nu_n} (\nu_n-\nu_j) / \nu_i \label{eq:gamma_simple} \\
    \Heat_{i,n} & = \Gamma_{i,\nu_n} h(\nu_n - \nu_i) \label{eq:Heat_simple}.
\end{align}    
The total rate coefficients can be found by summing each of these expressions
over all frequencies, $n=1,2,3,\ldots, n_{\nu}$. These equations are identical
to Equations (\ref{eq:Gamma_PhiPsi})--(\ref{eq:Heat_PhiPsi}) for the discrete
SED case, but are perhaps more intuitive.

For simplicity, our current treatment neglects a few physical processes that
are cosmological in origin, or simply do not rely on the radiation field
directly. These include cooling via free-free emission and hydrogen and helium
ionization due to helium recombination photons (which depend on the gas
kinetic temperature and electron density), and cosmological effects such as
Hubble cooling, Compton cooling off cosmic microwave background (CMB) photons,
and photo-ionization by Wien-tail CMB photons (which depend on kinetic
temperature, redshift, and the Hubble parameter).  

Two additional approximations are implicit in the remainder of this paper.
They are (1) the infinite speed-of-light approximation and (2) the on-the-spot
approximation (we use the case-B recombination coefficients in Equations
(\ref{eq:HIIRateEquation})--(\ref{eq:HeIIIRateEquation})). The former
approximation could be dubious for very bright sources in low-density media,
while the latter is generally not a good assumption, as discussed at length in
\citet{Cantalupo2011}. As a result, the \textit{absolute} accuracy of our
solutions is not guaranteed in regimes where careful treatment of the speed of
light and recombination photons is necessary, but this is acceptable since we
only care about the \textit{relative} differences among our solutions. The
optimized SEDs of Section \ref{sec:Results} will apply equally well to
simulations including more ionization and/or heating/cooling processes, so
long as they do not depend directly on the radiation field \citep[e.g.,
ionization of \HI\ and \HeI\ by helium recombination
photons;][]{Friedrich2012}.

\section{ASSESSING THE CONSEQUENCES OF DISCRETE RADIATION FIELDS} \label{sec:Consequences}
To quantify the differences between the ionization and temperature profiles
around sources with continuous and discrete SEDs, we will simulate two test
problems. First, the standard case of a $10^5$ K blackbody in a hydrogen-only
medium, and second, a power-law X-ray source in a medium consisting of both
hydrogen and helium.

\subsection{$10^5$ K Blackbody} \label{sec:BB}
The $10^5$ K blackbody problem has been studied extensively \citep[e.g., Test
Problem 2 in the Radiative Transfer Comparison Project;][hereafter
RT06]{Iliev2006} due to its simplicity, and perhaps also because the surface
temperatures of PopIII stars are expected to be $\sim 10^5 \ \mathrm{K} $
\citep{Schaerer2002}. We adopt nearly the identical setup as in RT06, i.e., a
uniform hydrogen-only medium with number density $\nH = 10^{-3} \
\mathrm{cm^{-3}}$, initial ionized fraction $\xHII = 1.2 \times 10^{-3}$,
initial temperature $T_0 = 10^2 \ \mathrm{K}$, and a $10^5 \ \mathrm{K}$
blackbody with an ionizing photon luminosity of $\dot{Q} = 5 \times 10^{48} \
\mathrm{s^{-1}}$. The only difference between our simulations and RT06 is that
we use a domain $L_{\mathrm{box}} = 10 \ \mathrm{kpc}$ in size, rather than
$L_{\mathrm{box}} = 6.6 \ \mathrm{kpc}$, to allow for a comparison of discrete
and continuous solutions at slightly larger radii. We evolve the simulations
for 500 Myr on a grid of 200 linearly spaced cells between $0.1 <
r/\mathrm{kpc} < 10$, ignoring the details of secondary ionization (i.e., all
photo-electron energy is deposited as heat).

\begin{figure*}[htbp]
\begin{center}
\includegraphics[width=0.98\textwidth]{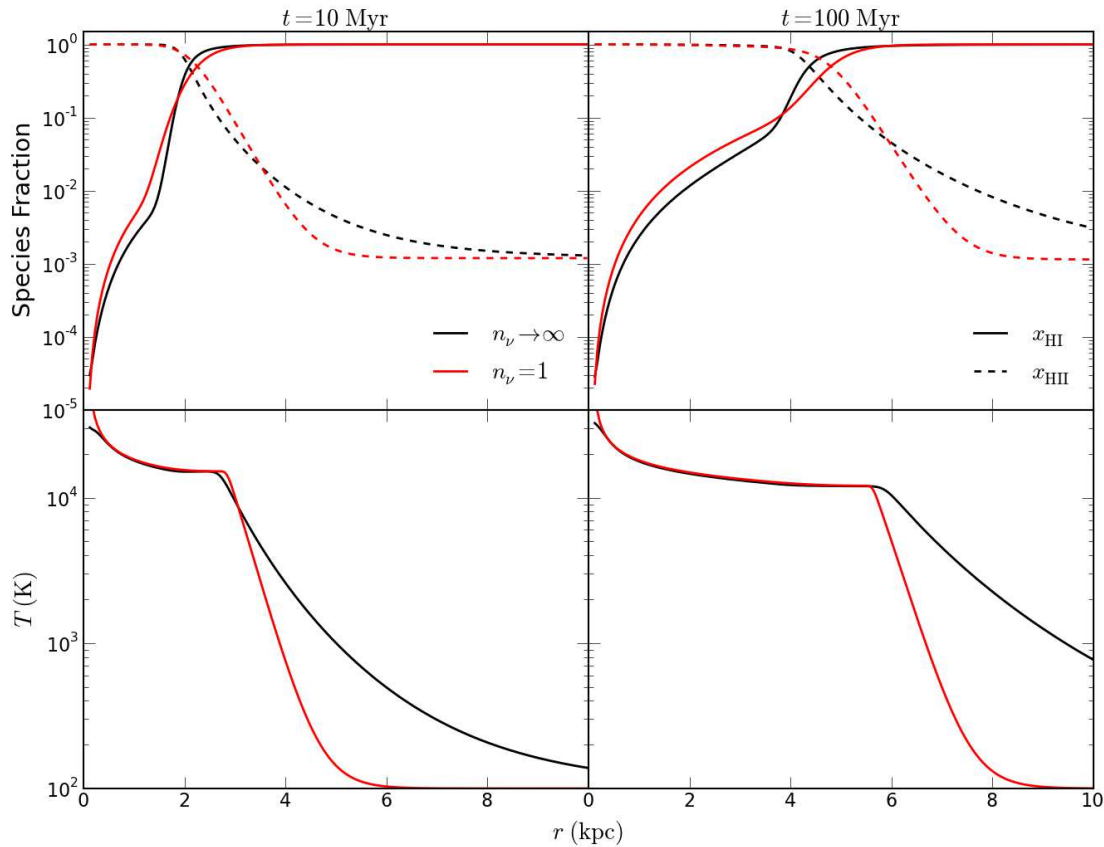}
\caption{Comparison of ionization (top) and temperature (bottom) profiles around a $10^5 \ \mathrm{K}$ blackbody source after 10 Myr (left) and 100 Myr (right) using continuous (black) and monochromatic (red) SEDs.  Solid lines in the top panels correspond to the neutral fraction ($\xHI$), while dashed lines  correspond to the ionized fraction ($\xHII$).  We apply these line color and line style conventions for all radial profiles presented in this paper.}
\label{fig:BB_DiscreteSpectrumTests}
\end{center}
\end{figure*}

In Figure \ref{fig:BB_DiscreteSpectrumTests}, we compare the ionization and
temperature profiles around two $10^5 \ \mathrm{K}$ `blackbody' sources of
constant ionizing photon luminosity $\dot{Q} = 5 \times 10^{48}s^{-1}$ --- one
a true blackbody emitter with a continuous SED spanning the range 13.6--100 eV
(black lines), and the other with a monochromatic SED at $h\nu_1 = 29.6$ eV,
the average energy of ionizing photons for this source (red lines). We can see
the same qualitative results that have been pointed out by previous authors,
namely, that monochromatic sources of radiation fail to ionize (top panels) and
heat (lower panels) gas at large radii as significantly as continuous sources,
since all photons are absorbed near a single characteristic column density,
representing the point where $\tau_{\nu_1} \approx 1$, i.e., $N_{\mathrm{char}}
\sim \sigma_{\nu_1}^{-1}$. The relative error in the position of the
ionization front, $\Delta r_{\mathrm{IF}}$, where $r_{\mathrm{IF}} \equiv
r(\xHI = \xHII = 0.5)$, is 8\% after 10 Myr, 10\% after 100 Myr, and 11\%
after 500 Myr. In the optically thin regime, the monochromatic spectrum
overestimates ionization by factors of two to three on average and up to an order of
magnitude at all times, though the latter effect is primarily because the
neutral fraction is a steeply declining function with decreasing radius, and
the I-fronts of the two solutions are offset. Outside the I-front, the
situation is more interesting as the gas is mostly neutral. After 100 Myr of
evolution, the ionized fraction outside the I-front is underestimated by a
factor of two on average, and by as much as a factor of six.

The temperature evolution, shown in the bottom panels of Figure
\ref{fig:BB_DiscreteSpectrumTests}, is significantly more troubling. The
monochromatic source captures the temperature well within the ionization front
where the gas is in photoionization equilibrium, but quickly diverges from the
continuous solution outside. Like the ionization profiles, discrepancies grow
with time. After 10 Myr of evolution, the monochromatic source underestimates
the temperature at large radii by a factor of two on average, and by a factor
of seven at the point of greatest discrepancy. After 100 (500) Myr, the
discrete solution underestimates the temperature by up to a factor of 17 (41).

If considering the heating and ionization around a single PopIII star, the
errors induced by monochromatic treatments may not be cause for concern upon
first inspection since PopIII stars are expected to live only a few Myr, and
we can see that errors are less significant at early times. However, the
intergalactic medium (IGM) is subject to the ionization and heating caused by
all sources, whose cumulative impact will be substantial even though the
ionization and heating caused by individual sources may be very small.
Globally, then, the IGM is insensitive to individual stellar lifetimes, and
instead evolves as it would if ionizing photons originated from a single, very
luminous, very long lived object.

This manner of thinking has already materialized in the realm of large volume
cosmological simulations, where `star particles' are generally as luminous as
one or more star clusters, and `galaxy particles' behave in a way that is
consistent with the integrated properties of an entire galactic stellar
population (and perhaps active nucleus). Such approximations are necessary
with limited spatial resolution, but more than adequate for studies of the
IGM. Over time though, errors in gas properties due to poor frequency
resolution will accrue, as it is the combined properties of all radiation
sources which affect IGM properties, however short-lived each individual
source may be.

\subsection{Power-Law X-Ray Source} \label{sec:PL}
To address the effects of discrete SEDs in environments where multiple
chemical species are important and large attenuating columns are possible, we
now turn our attention to a power-law X-ray source embedded in a 1 Mpc domain
consisting of hydrogen and helium, with a primordial helium abundance (by
mass) of $Y = 0.2477$.

Our selection of parameters for this problem is motivated by studies of
high-redshift quasars, and particularly their role in the epoch of
reionization \citep[e.g.,][]{Venkatesan2001}. X-rays have long mean free
paths, and as a result are capable of ionizing and heating gas on very large
($\sim$Mpc) scales. Large-scale heating is responsible for driving the
high-redshift all-sky 21 cm signal toward emission, and inducing fluctuations
in 21 cm power spectra on large angular scales (for a review of 21 cm
cosmology, see \citet{Furlanetto2006}). An early X-ray background may also be
important in interpreting the optical depth to electron scattering of the CMB
\citep[e.g.,][]{Ricotti2005,Shull2008}.

While supernovae and/or X-ray binaries could be important sources of hard
photons in the early universe, we assume the source of X-rays is persistent
--- an accreting SMBH with mass $M_{\bullet} = 10^6 M_{\odot}$ and radiative
efficiency of $\epsilon_{\bullet} = 10\%$, which leads to a bolometric
luminosity of $\mathcal{L}_{\mathrm{bol}} = \epsilon_{\bullet}
\mathcal{L}_{\mathrm{edd}} \simeq 1.26 \times 10^{43} \ \mathrm{erg \
s^{-1}}$. Here, $\mathcal{L}_{\mathrm{edd}} = 4\pi G M_{\bullet} m_p c /
\sigma_T$ is the Eddington luminosity, where $m_p$ is the proton mass and
$\sigma_T$ the Thomson cross-section. The mass (and thus luminosity) of the
SMBH is allowed to grow as it accretes,
\begin{equation}
    M_{\bullet}(t) = M_{\bullet}(0) \mathrm{exp}\left[\frac{1-\epsilon_{\bullet}}{\epsilon_{\bullet}}\left(\frac{t}{t_{\mathrm{edd}}}\right)\right] ,
\end{equation}    
where $t_{\mathrm{edd}} = 0.45$ Gyr is the $e$-folding timescale for SMBH
growth (an Eddington, or Salpeter time). The SED is taken to be a power law of
the form
\begin{equation}
    I_{\nu} \propto \left(\frac{h\nu}{\mathrm{keV}} \right)^{1 - \alpha},
\end{equation}    
where $\alpha$ is the spectral index. We adopt $\alpha = 1.5$, over the energy
range $10^2$-$10^4$ eV. The surrounding medium has a constant mass density of
$\rho = 5.4 \times 10^{-28} \ \mathrm{g \ cm^{-3}}$ (cosmic mean at redshift
$z = 10$), initial ionized fractions $\xHII = \xHeII = 10^{-4}$, $\xHeIII =
0$, and initial temperature $T_0 = 10^2 \ \mathrm{K}$. The domain for this
problem is divided into 400 cells linearly spaced between $0.01 <
r/\mathrm{Mpc} < 1$, and is evolved for $\epsilon_{\bullet} t_{\mathrm{edd}} =
45$ Myr.

\begin{figure}[htbp]
\begin{center}
\includegraphics[width=0.48\textwidth]{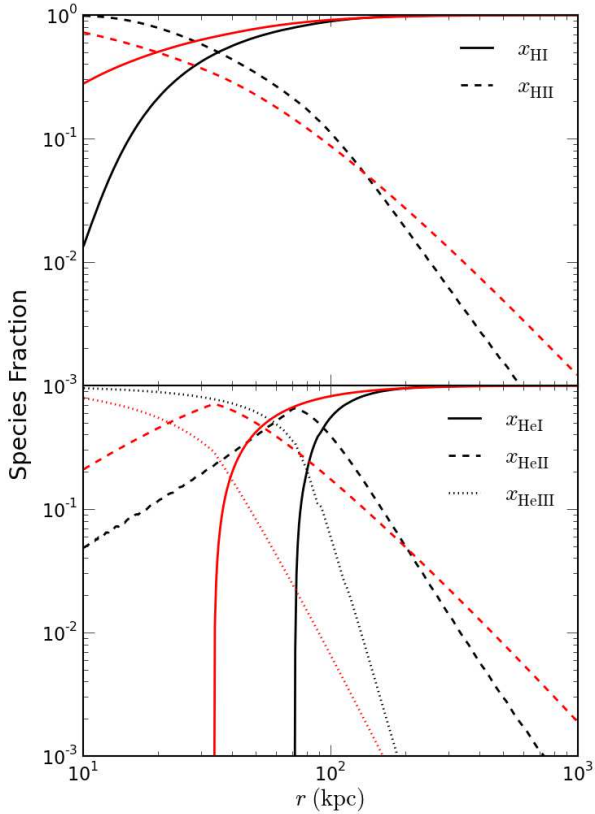}
\caption{Comparison of hydrogen (top) and helium (bottom) ionization profiles around an $\alpha = 1.5$ power-law X-ray source after 45 Myr using continuous (black) and monochromatic (red) SEDs.}
\label{fig:Xray_DiscreteSpectrumTests_spfrac}
\end{center}
\end{figure} 

\begin{figure}[htbp]
\begin{center}
\includegraphics[width=0.48\textwidth]{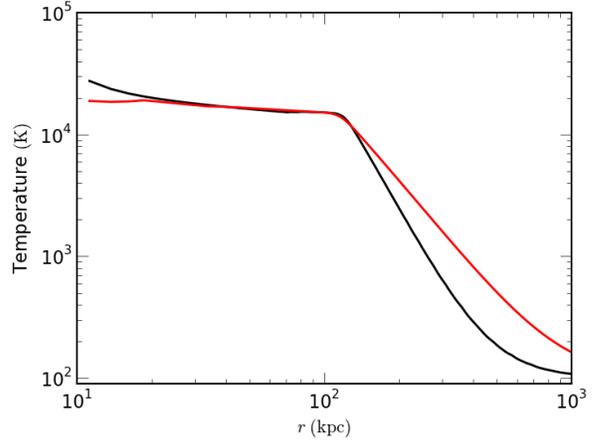}
\caption{Comparison of temperature profiles around an $\alpha = 1.5$ power-law X-ray source after 45 Myr using continuous (black) and monochromatic (red) SEDs.}
\label{fig:Xray_DiscreteSpectrumTests_T}
\end{center}
\end{figure} 

In Figure \ref{fig:Xray_DiscreteSpectrumTests_spfrac}, we compare the hydrogen
and helium ionization profiles for two X-ray sources having the same
bolometric luminosity. One, a continuous power-law source as described above,
and the other a monochromatic source of $0.5$ keV photons (a fiducial
monochromatic emission energy). The monochromatic source underestimates the
radii of both the hydrogen and helium ionization fronts by a factor of $\sim
2.3$, and overestimates the hydrogen neutral fraction on average by a factor
of three, and at most by a factor of 20 within the hydrogen I-front. The same
general picture applies to helium, where errors in the neutral helium fraction
are enormous since the \HeI-\HeII\ I-front is very sharp (as it was for
hydrogen in the previous section), and $\xHeII$ and $\xHeIII$ are in error by
factors of 2--20 depending on radius.

Errors in the temperature profile are less extreme, as shown in Figure
\ref{fig:Xray_DiscreteSpectrumTests_T}. On small scales, the monochromatic
source captures the temperature quite well, but at large radii, the
monochromatic source overestimates temperatures by a factor of two on average.

The disparity in the magnitude of ionization and temperature errors is a
reflection of the strong frequency dependence of the bound--free absorption
coefficients. Photo-ionization of hydrogen or helium by $0.5$ keV photons is
rare, but when it does occur, at least $\sim 90\%$ of the original photon
energy is left to be deposited mostly as heat, unless the free electron
density is very low. Because the ionization of hydrogen and helium by the
monochromatic source is very inaccurate, errors in the free electron density
will substantially alter the amount of secondary electron energy deposited as
heat, rather than further ionization.

The consequences of miscalculating ionization and heating could affect efforts
to model and interpret current and future 21 cm measurements, since the
primary 21 cm observable, the differential brightness temperature (\dTb),
depends on the hydrogen neutral fraction, UV radiation field, electron
density, and the gas kinetic temperature ($T_K$) \citep{Furlanetto2006}.
Neglecting the presence of a Ly$\alpha$ background, the scaling
\begin{equation}
   \delta T_b \propto T_K^{0.4} (1 + \delta) (1 + z)^{-1/2} \times \left\{
     \begin{array}{lr}
       \xHI \nel  &, \nel \gg \nHI \\
       \xHI^2  &, \nel \ll \nHI 
     \end{array}
   \right.
\end{equation}
holds approximately in regimes where $T_{\mathrm{CMB}} \ll T_K \lesssim 10^4$ K. 

In the immediate vicinity of radiation sources where gas is entirely ionized,
$\delta T_b \rightarrow 0$ due to the leading $\xHI$ term, but at large radii
where the ionizing flux is weaker, the \dTb\ signatures of stars and quasars
could vary significantly solely due to miscalculations of $\xHI$, $\nel$, and
$T_K$. The above scalings have especially strong consequences for gas within a
few Mpc of strong X-ray sources, where hydrogen is weakly ionized,
temperatures are of order $10^2$-$10^3$ K, and the free electron density is
enhanced due to efficient ionization of helium by the hard radiation field. In
the earliest stages of reionization where $T_K < T_{\mathrm{CMB}}(z)$ and the
Ly$\alpha$ background is important, errors in $\xHI$, $\nel$, and $T_K$
will lead to errors in $\delta T_b$ as well, though in a less straightforward
way, since the spin temperature, $T_S$, must be computed carefully.

\section{Optimization Strategy} \label{sec:Methods}
To avoid errors of the sort described in the previous section, we have
developed a technique for optimally constructing discrete SEDs that preserves
the ionization and heating properties of their continuous counterparts.
Although ray-tracing algorithms are capable of tabulating the relevant
ionization and heating quantities (Equations (\ref{eq:PHI}) and (\ref{eq:PSI})),
few codes have taken advantage of this, and have instead cast monochromatic
rays \citep[e.g., state of the art reionization simulations with $n_{\nu} =
5$;][]{Trac2008}. Monte Carlo codes \citep[e.g., CRASH;][]{Maselli2003} have
been used to simulate reionization with $n_{\nu} \geq 20$ multi-frequency
photon packets \citep{Ciardi2011}, though such a large number of frequencies
may be computationally debilitating for some algorithms, or unnecessary
depending on the problem of interest.

Even when the algorithm of choice is compatible with propagating continuous
radiation fields via tabulation of Equations (\ref{eq:PHI}) and
(\ref{eq:PSI}), it may not be computationally advantageous. The overhead alone
can in fact be substantial, particularly in the case of source-dependent SEDs
--- for example, the SED of a stellar population as a function of age, or BH
accretion spectra that vary with mass or luminosity. Such situations would
require a separate lookup table for Equations (\ref{eq:PHI}) and
(\ref{eq:PSI}) at each age/mass/luminosity of interest for a given radiation
source. In addition, there are algorithms for which propagating continuous
radiation fields in large volumes become completely intractable, yet large
volumes are a necessity for the science questions of interest (e.g.,
reionization). For more discussion on these issues, see Section
\ref{sec:Discussion}.

As introduced in Section \ref{sec:Framework}, our optimization strategy relies
on the fact that the SED of a radiation source appears only in the quantities
$\Phi_i$ and $\Psi_i$ (see Equations (\ref{eq:PHI}) and (\ref{eq:PSI})). If we
can construct a discrete SED that reproduces the values of $\Phi_i$ and
$\Psi_i$ to a high degree of accuracy over a column density interval of
interest, then the discrete radiation field is indistinguishable from its
continuous counterpart, and we have successfully preserved the true radiative
properties of the source.

For sources with discrete SEDs, Equations (\ref{eq:PHI}) and (\ref{eq:PSI}) become
\begin{align}
    \Phi_i^{\prime} (\tau_{\nu_n}) & \equiv \sum_{n = 1}^{n_\nu} \frac{I_n}{h\nu_n} e^{-\tau_{\nu_n}} \label{eq:PHI_prime} \\
    \Psi_i^{\prime} (\tau_{\nu_n}) & \equiv \sum_{n = 1}^{n_\nu} I_n e^{-\tau_{\nu_n}} \label{eq:PSI_prime} ,
\end{align}
where we have used primes to indicate that these quantities are computed by
direct summation over $n = 1, 2, \ldots, n_\nu$ frequencies, rather than by a
continuous integral.

Ensuring that $\Phi_i = \Phi_i^{\prime}$ and $\Psi_i = \Psi_i^{\prime}$ is a
minimization problem of dimensionality $2n_{\nu}$, since each additional
frequency bin lends two degrees of freedom --- its frequency ($\nu_n$), and
the fraction of the bolometric luminosity assigned to that frequency ($I_n$).
Our goal is to minimize the difference between continuous and discrete
solutions, i.e.,
\begin{align}
    \Phi_i - \Phi_i^{\prime} & = 0 \nonumber \\
    \Psi_i - \Psi_i^{\prime} & = 0 . \label{eq:DumbOptimization}
\end{align}
These functions span several orders of magnitude over a broad range in column
density, making it more practical to seek solutions to
\begin{align}
    \mathrm{log}\left( \frac{\Phi_i}{\Phi_i^{\prime}}\right) & = 0 \nonumber \\ \mathrm{log}\left( \frac{\Psi_i}{\Psi_i^{\prime}}\right) & = 0  \label{eq:logMinimize}
\end{align} 
which place equal emphasis on all column densities. Preserving the high column
density behavior of $\Phi_i$ and $\Psi_i$ is especially important for very
luminous sources and/or environments with dense clumps in the immediate
vicinity of the source, since the actual photoionization and heating rates are
a combination of $\Phi_i$, $\Psi_i$, and the normalization factor $A_i \propto
L_{\mathrm{bol}} / r^2$. 

For a given $n_{\nu}$ and source SED, we solve Equation (\ref{eq:logMinimize})
using the optimization technique Simulated Annealing \citep{Kirkpatrick1983,
Cerny1985}, which traverses our $2n_{\nu}$ dimensional parameter space in
search of the frequency--normalization pairs $(\nu_n, I_n)$ that best
reproduce the values of $\Phi_i$ and $\Psi_i$. We leave a more detailed
description of the algorithm and our implementation of it to the Appendix.

\section{RESULTS} \label{sec:Results}
\subsection{Optimal Discrete SEDs} \label{sec:OptimalSEDs}
We have obtained optimal SEDs for a $10^5$ K blackbody emitting in the range
$13.6$-$100$ eV, and an $\alpha = 1.5$ power-law X-ray source with emission
spanning the interval $10^2$-$10^4$ eV. In each case, we set the upper column
density limit for our optimization to be the column density of a fully neutral
medium, i.e., $\NHI^{\mathrm{max}} = \nH \Lbox$ and $\NHeI^{\mathrm{max}}
= \nHe \Lbox$, where we use $\Lbox$ to denote the size of the domain, as in
RT06. For the $10^5$ K blackbody simulations, this works out to be
$\NHI^{\mathrm{max}} = 3.1 \times 10^{19} \ \mathrm{cm^{-2}}$, and for the
power-law X-ray simulations, $\NHI^{\mathrm{max}} \simeq \times 10^{22} \
\mathrm{cm^{-2}}$ and $\NHeI^{\mathrm{max}} \simeq \times 10^{21} \
\mathrm{cm^{-2}}$. For cosmological simulations with periodic boundary
conditions, the upper column density limits would need to be chosen based on a
maximum length scale of interest, or for radiative feedback focused
simulations, by the column density of the densest objects of interest (damped
Ly$\alpha$ systems, for example). Such choices are already made in
ray-tracing calculations to limit computational expense. Generally, rays are
terminated once the emission has been attenuated by a large factor.

The only situation in which we do not evaluate the full cost function is
$n_{\nu} = 1$, where we instead optimize for the optically thin regime alone
(i.e., only the first term of Equation \ref{eq:Cost}), where $\Phi_i$ and
$\Psi_i$ are $\sim$ constant with column density. In this case, the optimal
solutions are simply those that preserve the bolometric luminosity of the
source and the total number of ionizing photons, and can be verified
analytically (Equations (\ref{eq:PHI}) and (\ref{eq:PSI})). For the case of a
hydrogen and helium medium, we have found that neglecting \HeII\ opacities
mitigates the computational cost of the computation while resulting in no
appreciable changes in our optimal SEDs and thus negligible changes in
$\Phi^{\prime}$ and $\Psi^{\prime}$. The main results are summarized in
Figures \ref{fig:BB_hists} and \ref{fig:PL_hists} and Tables
\ref{tab:OptimalBBspectra} and \ref{tab:OptimalPLspectra}, all results derived
from $K = 2\times 10^4$ and $K = 10^4$ Monte-Carlo trials, for the $10^5$ K
blackbody and $\alpha = 1.5$ power-law source, respectively.

\begin{deluxetable}{ccccc}
\tablewidth{0pt}
\tabletypesize{\small}
\tablecaption{Optimal SEDs for $10^5$ K Blackbody Sources}
\tablecolumns{6}
\tablehead{\colhead{$n_{\nu}$} & \colhead{$n = 1$} & \colhead{$n = 2$} & \colhead{$n = 3$} & \colhead{$n = 4$}}
\startdata
1 & $(29.61, 0.89)$ & $\ldots$ & $\ldots$ & $\ldots$ \\
2 & $(27.93, 0.68)$ & $(62.04, 0.21)$ & $\ldots$ & $\ldots$ \\
3 & $(20.58, 0.39)$ & $(40.75, 0.39)$ & $(69.23, 0.11)$ & $\ldots$ \\
4 & $(17.98, 0.23)$ & $(31.15, 0.36)$ & $(49.09, 0.24)$ & $(76.98, 0.06)$ 
\enddata
\tablecomments{Each entry is the $(h\nu_n, I_n)$ pair for bin $n$.  Energies are in units of eV, and normalizations are expressed as fraction of the bolometric luminosity.}
\label{tab:OptimalBBspectra}
\end{deluxetable}

\begin{deluxetable}{ccccc}
\tablewidth{0pt}
\tabletypesize{\small}
\tablecaption{Optimal SEDs for $\alpha = 1.5$ Power-Law X-ray Sources}
\tablecolumns{6}
\tablehead{\colhead{\nnu} & \colhead{$n = 1$} & \colhead{$n = 2$} & \colhead{$n = 3$} & \colhead{$n = 4$}}
\startdata
1 & $(999.98, 1.00)$ & $\ldots$ & $\ldots$ & $\ldots$ \\
2 & $(255.87, 0.17)$ & $(2553.6, 0.83)$ & $\ldots$ & $\ldots$ \\
3 & $(171.93, 0.08)$ & $(518.22, 0.14)$ & $(3098.5, 0.78)$ & $\ldots$ \\
4 & $(146.11, 0.05)$ & $(307.30, 0.07)$ & $(704.56, 0.14)$ & $(3564.2, 0.73)$ 
\enddata
\tablecomments{Same as Table \ref{tab:OptimalBBspectra} but for an $\alpha = 1.5$ power-law X-ray source.}
\label{tab:OptimalPLspectra}
\end{deluxetable}

\begin{figure}[ht]
\centering
\subfigure[]{
\includegraphics[width=0.48\textwidth]{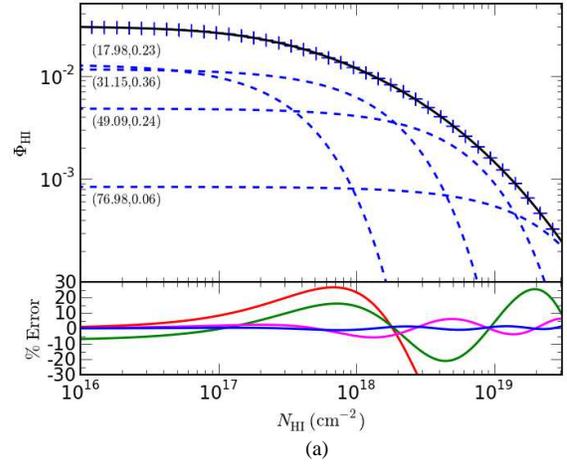}
\label{fig:BB_Phi}
}
\subfigure[]{
\includegraphics[width=0.48\textwidth]{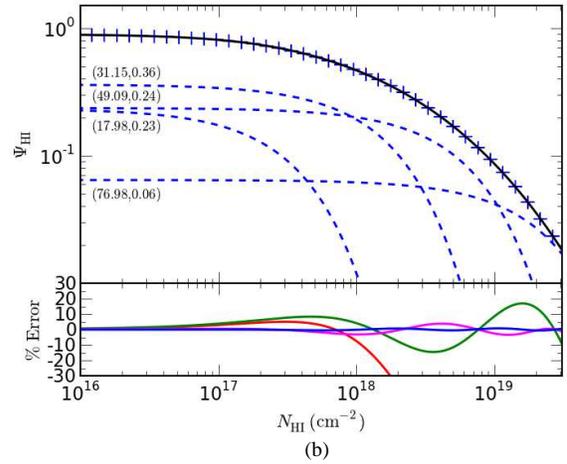}
\label{fig:BB_Psi}
}
\caption[]{Top Panels: Comparison of $\PhiHI$ and $\PhiHI^{\prime}$ (a) and $\PsiHI$ and $\PsiHI^{\prime}$ (b) as a function of \HI\ column density for a $10^5 \ \mathrm{K}$ blackbody, showing the numerically computed continuous integral (solid black), best-fit composite four-bin discrete sum (blue crosses), and the contribution from each individual discrete frequency bin (dashed blue). Annotations represent the $(h\nu_n, I_n)$ pairs for each frequency group, drawn from Table \ref{tab:OptimalBBspectra}. Bottom Panels: Percent error between discrete and continuous solutions.  The solid blue line is the error for the four-bin optimal solution, while the errors induced by three-, two-, and one-bin solutions are shown in magenta, green, and red, respectively.}
\label{fig:BB_PhiPsi}
\end{figure}

\begin{figure}[ht]
\centering
\subfigure[]{
\includegraphics[width=0.48\textwidth]{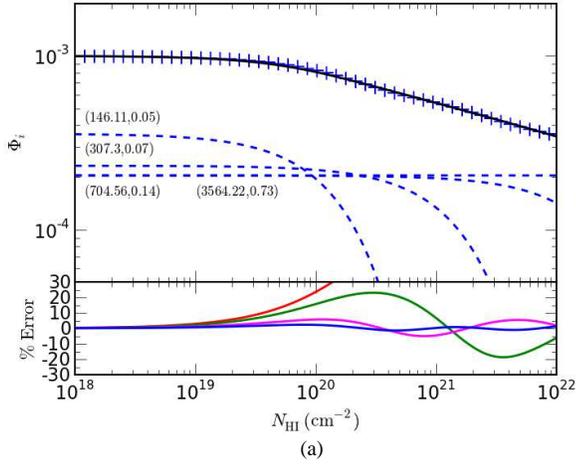}
\label{fig:PL_Phi}
}
\subfigure[]{
\includegraphics[width=0.48\textwidth]{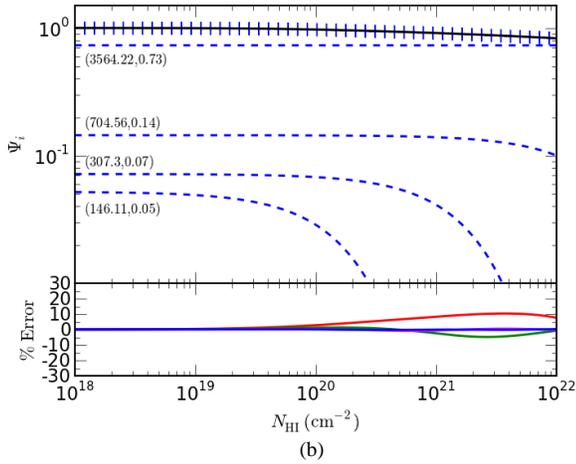}
\label{fig:PL_Psi}
}
\caption[]{Same as Figure \ref{fig:BB_PhiPsi} but for an $\alpha = 1.5$ power-law X-ray source.}
\label{fig:PL_Phi_Psi}
\end{figure}

From Tables \ref{tab:OptimalBBspectra} and \ref{tab:OptimalPLspectra}, it is
clear that the optimal emission frequencies for both sources are not evenly
spaced above the hydrogen or helium ionization thresholds, either in linear or
log-space. In each case, the addition of a new frequency bin leads to a
decrease in both the emission frequency and normalization of all other bins.
This signifies (1) the efficacy with which high energy photons photoionize and
photoheat gas at large column densities (a regime inaccessible to lower
energy photons which become optically thick at small columns), and (2) the
increase in excess electron kinetic energy available for further ionization
and heating with increasing photon energy. The former effect is most important
for the blackbody source, which we can see in Figure \ref{fig:BB_PhiPsi}. Not
surprisingly, it is the lowest energy photons ($h\nu_1 = 17.98$ eV) in the
$n_{\nu} = 4$ spectrum that are responsible for the ionization (through
$\Phi$) in the optically thin regime, while successively higher frequency bins
become the primary agents of ionization as we move to higher column densities.
The same trend does not hold completely in Figure \ref{fig:BB_Psi}, as in this
case it is the second and third energy bins that provide the bulk of the
heating (through $\Psi$) at low column densities.

\begin{figure}[ht]
\centering
\subfigure[]{
\includegraphics[width=0.48\textwidth]{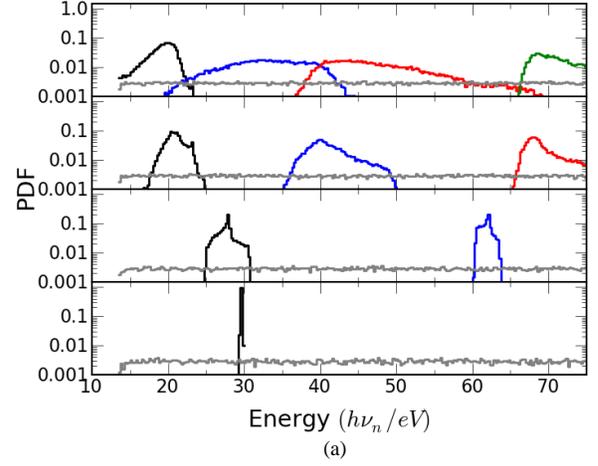}
\label{fig:BB_hist_E}
}
\subfigure[]{
\includegraphics[width=0.48\textwidth]{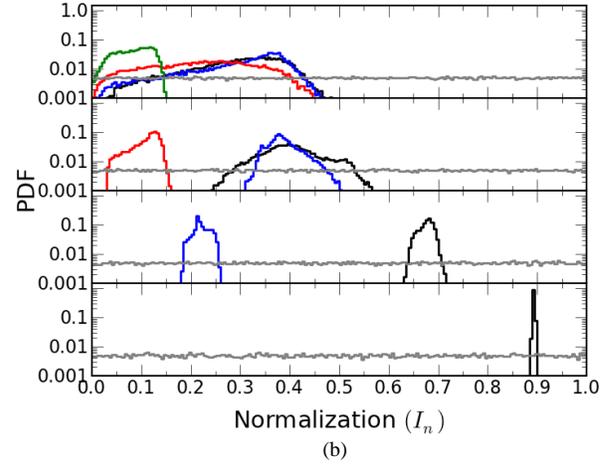}
\label{fig:BB_hist_F}
}
\caption[]{Emission energy (a) and normalization (b) probability distribution functions (PDFs) of optimized discrete $10^5$ K blackbody spectrum using $n_{\nu} = 1,2,3,4$ (from bottom to top). In each panel, the gray histogram denotes the initial guesses for all Monte-Carlo trials, and the black, blue, red, and green histograms show the end point for the first, second, third, and fourth bins, respectively (ordered by increasing emission frequency).}
\label{fig:BB_hists}
\end{figure}

\begin{figure}[ht]
\centering
\subfigure[]{
\includegraphics[width=0.48\textwidth]{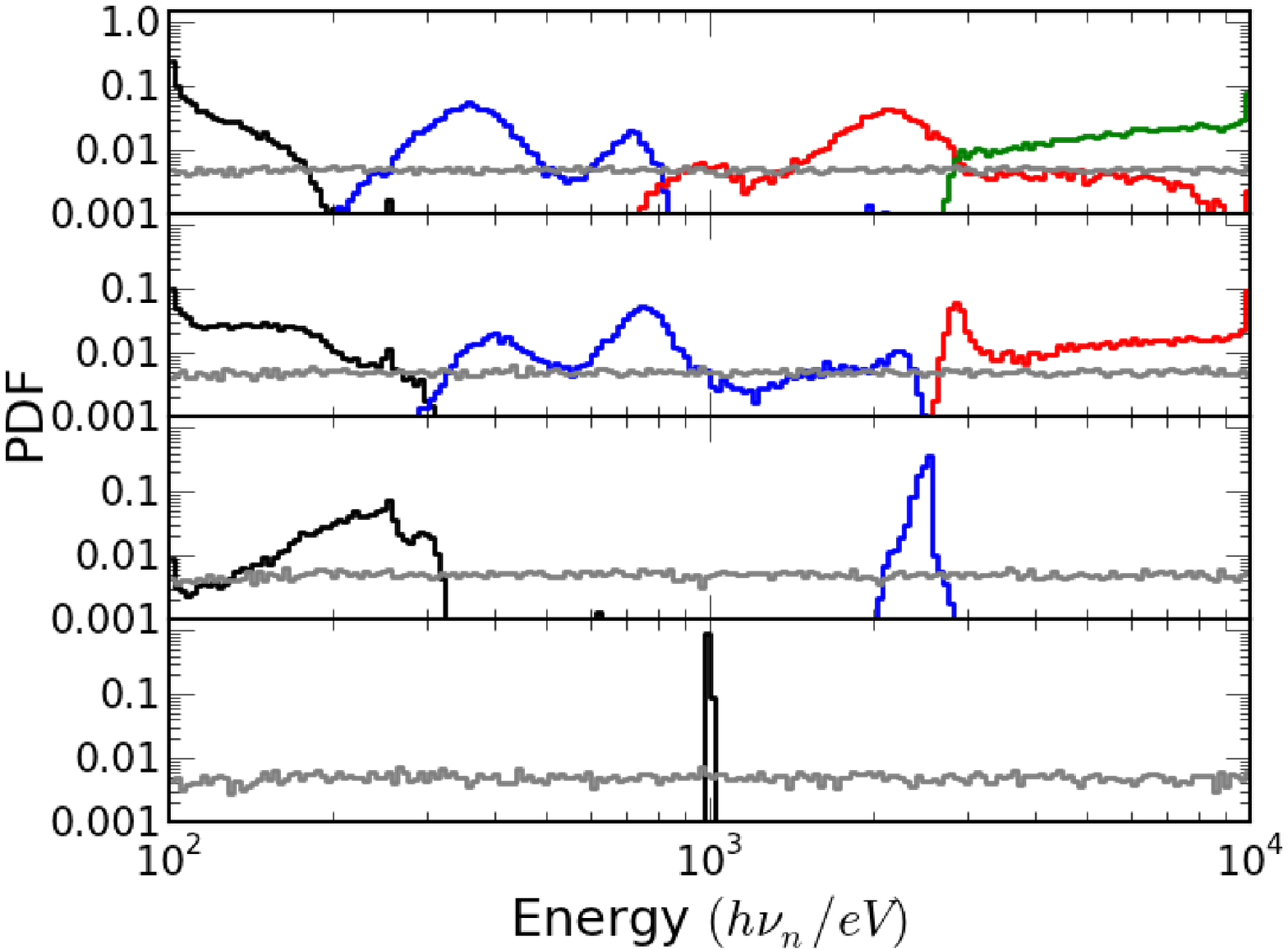}
\label{fig:PL_hist_E}
}
\subfigure[]{
\includegraphics[width=0.48\textwidth]{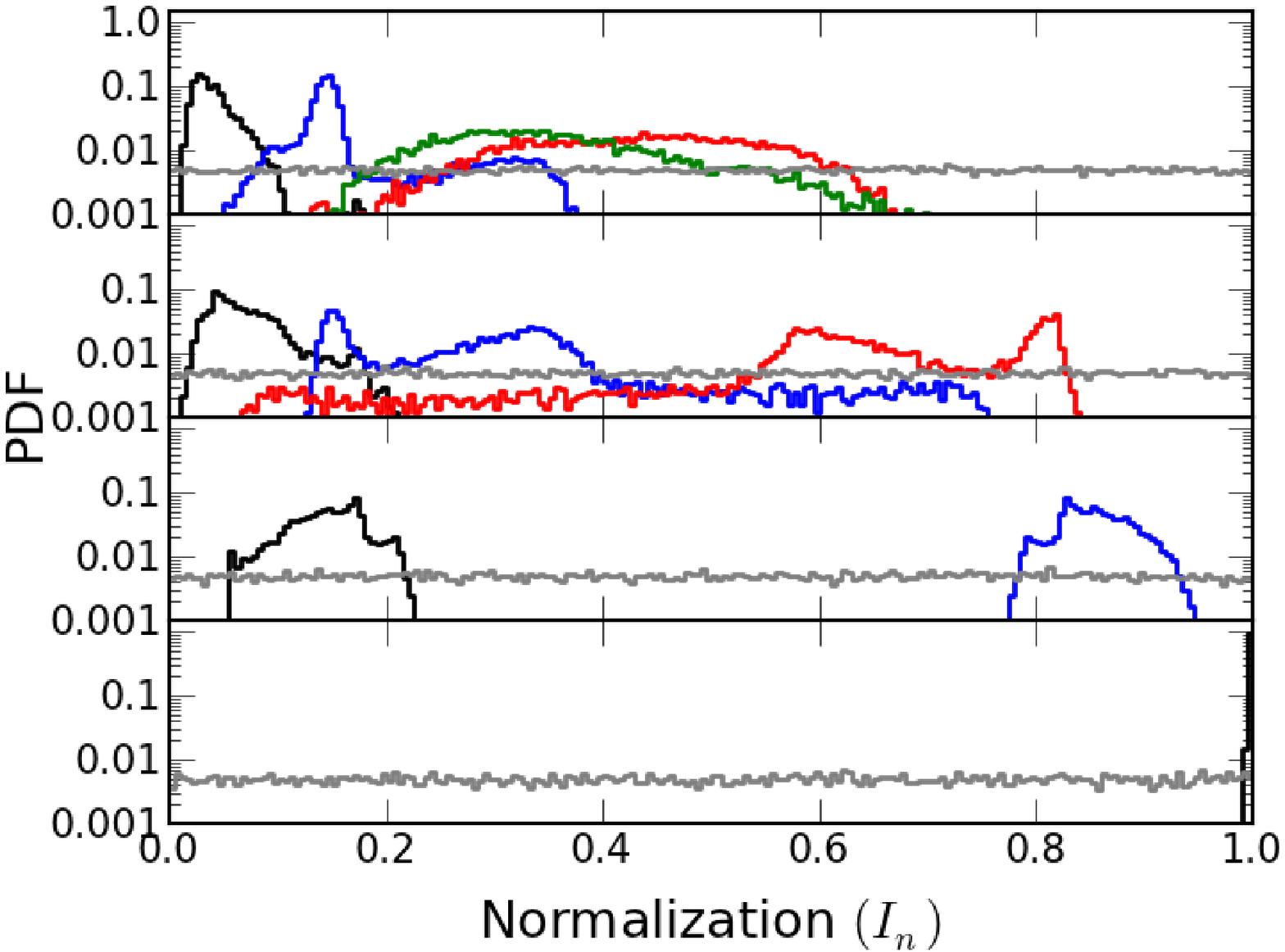}
\label{fig:PL_hist_F}
}
\caption[]{Same as Figure \ref{fig:BB_hists} but for an $\alpha = 1.5$ power-law X-ray source.}
\label{fig:PL_hists}
\end{figure}

For the X-ray source, the second effect dominates, as the optical depth at any
column density is small for most photons considered ($10^2 < h\nu < 10^4$ eV)
over the entire domain. As shown in Figure \ref{fig:PL_Phi_Psi}, the photons
responsible for the majority of the heating (through $\Psi$) over all column
densities are those in the highest energy bin, the same photons which are the
least effective at ionization. The trends and errors of Figure
\ref{fig:PL_Phi_Psi} are the same for $\Phi_i$ and $\Psi$ as a function of
helium column density.

In Figures \ref{fig:BB_hists} and \ref{fig:PL_hists}, we show the probability
distribution functions (PDFs) for the position and normalization of the
optimal SED frequency bins obtained (drawn from Tables
\ref{tab:OptimalBBspectra} and \ref{tab:OptimalPLspectra}). Solutions are less
tightly constrained as $n_{\nu}$ is increased, as evidenced by a broadening in
the distributions of frequency and normalization for each bin. This behavior
is expected, given that each new bin contributes to the magnitude of $\Phi$
and $\Psi$ in some region of column density space previously occupied by one
or more other frequencies.

Holding $I_n$ constant, a decrease in $\nu_n$ will cause a negative vertical
shift in the contribution of bin $n$ to the magnitude of $\Phi$, for example,
but will simultaneously add power at larger column densities, since the
turnover point for bin $n$ occurs at $N_{\mathrm{char}} \sim
\sigma_{\nu_n}^{-1}$, and $\sigma_{\nu_n} \sim \nu^{-3}$. To avoid an increase
in $f$, the power lost at small column densities has to be compensated for,
either by a decrease in $\nu_{n-1}$, or an increase in $I_{n-1}$, where $n-1$
denotes the bin with frequency $\nu_{n-1} < \nu_n$. As a result, there are
degeneracies between all bins, and the magnitude of the degeneracy is greatest
for bins positioned closest in frequency-space. In order to tighten the PDFs
for each optimal frequency bin, one or more terms would need to be added to
$f$, in order to assign preference to one set of bins over another. For our
purposes, any SED that minimizes $f$ is just as good as any other, but
additional terms in the cost function are certainly justifiable in the case of
a ray-tracing calculation, where higher emission frequencies increase the
computational cost of a calculation since their mean free paths are long.
Adding a term to $f$ that scales with $\nu_n$ would encourage optimal SEDs
with the smallest emission frequencies possible, for example.

Optimization for $n_{\nu} > 4$ is certainly possible, though unnecessary in
our case. At a given frequency, the transition from optically thin $\tau = 0$
to optically thick ($\tau \gtrsim 1$) in the functions $\Phi$ and $\Psi$
occurs over an order of magnitude in column density (by definition, see
Equation (\ref{eq:OpticalDepth})). For both SEDs we have investigated, the
column density regime of interest spans fewer than four orders of magnitude,
motivating our choice of $1 \leq n_{\nu} \leq 4$. We have performed
optimizations with $n_{\nu} > 4$, but the addition of each additional bin when
$n_{\nu} > \mathrm{log}_{10}(N_{\mathrm{max}}/N_{\mathrm{min}})$ reduces the
error between $\Phi$ and $\Phi^{\prime}$, and $\Psi$ and $\Psi^{\prime}$ much
less significantly than additional bins when $n_{\nu} \leq
\mathrm{log}_{10}(N_{\mathrm{max}}/N_{\mathrm{min}})$. For a given $n_{\nu}$,
increasing $N_{\mathrm{max}}$ will simply increase $\mathrm{max}|\Phi -
\Phi^{\prime}|$ and $\mathrm{max}|\Psi - \Psi^{\prime}|$.

\subsection{Confirmation with One-dimensional Calculations}
To verify the solutions of the previous section, we ran simulations identical
to those of Section \ref{sec:Consequences} but with our optimal discrete SEDs.
We compute $\Gamma_i$, $\gamma_i$, and $\Heat_i$ via Equations
(\ref{eq:Gamma_simple})--(\ref{eq:Heat_simple}) ``on-the-fly,'' rather than
generating lookup tables of $\Phi_i$ and $\Psi_i$. As expected, accurate
preservation of the quantities $\Phi_i$ and $\Psi_i$ over the column density
ranges of interest renders ionization and temperature profiles around sources
of discrete radiation indistinguishable from their continuous counterparts.

\begin{figure}[htbp]
\begin{center}
\includegraphics[width=0.48\textwidth]{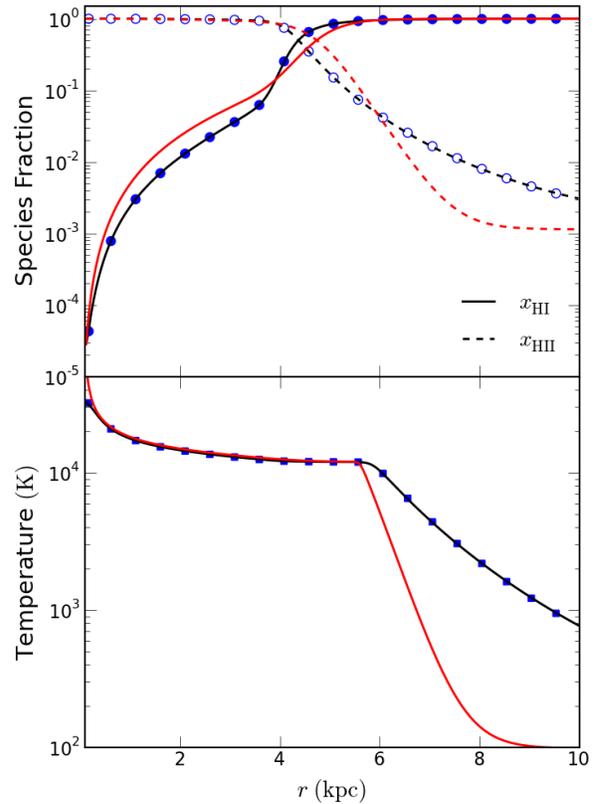}
\caption{Comparison of ionization (top) and temperature (bottom)
profiles around a $10^5$ K blackbody source after 100 Myr showing the solutions obtained using continuous (black), monochromatic (red), and optimal four-bin discrete (blue circles/squares) SEDs.}
\label{fig:BB_DiscreteSpectrumTests_opt}
\end{center}
\end{figure}

In Figure \ref{fig:BB_DiscreteSpectrumTests_opt}, we compare ionization and
heating around a $10^5$ K blackbody after 100 Myr of evolution as in Section
\ref{sec:Consequences}, showing the solution obtained with our optimal
monochromatic (red) and four-bin (blue) SEDs. The continuous and four-bin
solutions are indistinguishable.

\begin{figure}[ht]
\centering
\subfigure[]{
\includegraphics[width=0.48\textwidth]{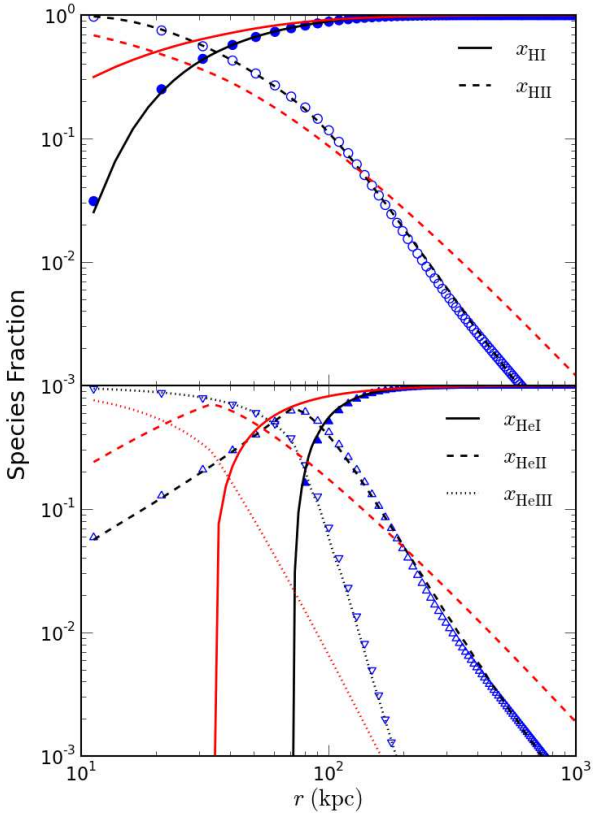}
\label{fig:Xray_DiscreteSpectrumTests_opt1}
}
\subfigure[]{
\includegraphics[width=0.48\textwidth]{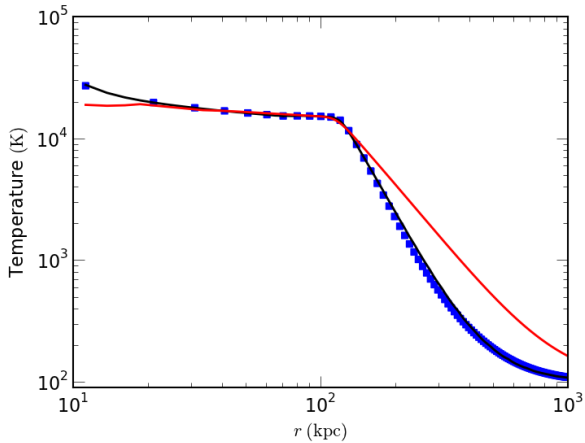}
\label{fig:Xray_DiscreteSpectrumTests_opt2}
}
\caption[]{Comparison of hydrogen and helium ionization (a) and temperature profiles (b) around a power-law X-ray source after 50 Myr showing the solutions obtained using continuous (black) and optimal four-bin discrete (blue symbols)
SEDs.}
\label{fig:Xray_DiscreteSpectrumTests_opt}
\end{figure}

In Figure \ref{fig:Xray_DiscreteSpectrumTests_opt}, we perform the same
analysis for the $\alpha = 1.5$ power-law simulations. Our optimal four-bin
SED reproduces the hydrogen and helium ionization profiles (and thus electron
density) and temperature of a continuous SED to high precision. The most
noticeable errors are in the hydrogen neutral fraction within the hydrogen
ionization front, where errors between four-bin and continuous solutions are
still only $\sim 1\%$. Errors in $\xHeIII$ are negligible, justifying our
neglect of $\NHeII$ in the optimization process.

It should be noted that our optimal monochromatic SED for the X-ray source
performs even more poorly than the fiducial $0.5$ keV SED. This signifies a
general problem with monochromatic emission for any spectrum with a hard
component. Whereas the monochromatic optimization ($\tau_{\nu} = 0$) works
quite well in the $10^5$ K blackbody case since hydrogen absorbs UV photons
readily, X-rays are not so readily absorbed by hydrogen and/or helium. As a
result, the characteristic column density where most 1 keV photons are
absorbed lies outside of our domain, leading to severe under-ionization (of
all species) and under-heating. The reason the $0.5$ keV SED works better is
because its characteristic absorption column is smaller, lying within our
domain. We have experimented with relaxing the optically thin requirement for
monochromatic optimization, and find that it is equally difficult to preserve
ionization and heating profiles with emission at a single frequency.

\subsection{Three-dimensional Radiation-hydrodynamic Simulations with \textit{Enzo}}
To study the impact of spectral discretization in a more complex setting, we
ran RT06 test problem 2 with hydrodynamics, as well as two fully
three-dimensional cosmological radiation-hydrodynamic simulations similar to
those of \citet{Abel2007} and \citet{Alvarez2009}, both with the \textit{Enzo}
code \citep{Bryan1997,Oshea2004}\footnote{Revision \texttt{f4a8b5f5e6c5},
modified to form only one star and use optimal SEDs.}. All analysis was
performed with \textit{yt} \citep{Turk2011}.

\begin{figure}[htbp]
\begin{center}
\includegraphics[width=0.48\textwidth]{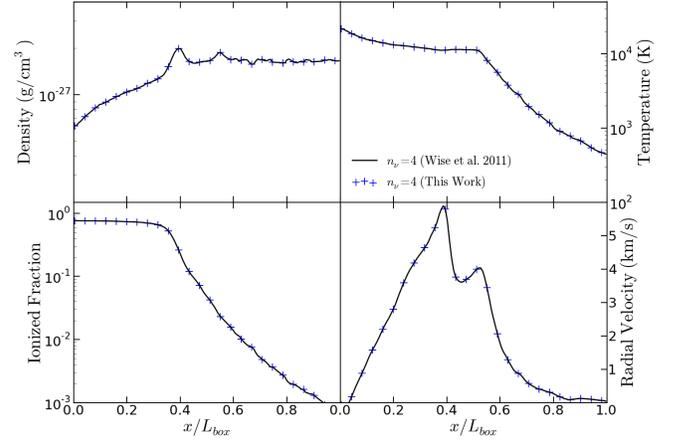}
\caption{Comparison of the four-bin solutions of \citet{Wise2011} (black) and our own (blue crosses) in a radiation-hydrodynamic simulation using the \textit{Enzo} code.  The setup is the same as in RT06 Test Problem 2, except hydrodynamics is included.}
\label{fig:wise_mirocha_4bin}
\end{center}
\end{figure}

The results of the RT06 radiation-hydrodynamic test problem are shown in
Figure \ref{fig:wise_mirocha_4bin}, where we compare the solutions obtained
using the four-bin SED employed by \citet{Wise2011} in addition to our own
(Table \ref{tab:OptimalBBspectra}). The solutions are indistinguishable, which
is expected given the relatively small range of column density explored in
this problem.

The cosmological simulations follow the formation of a $100 M_{\odot}$ PopIII
star, its brief $2.7$ Myr lifetime in which it emits $1.2 \times 10^{50}$
ionizing photons per second, and the X-ray emission resulting from accretion
onto a remnant BH assumed to form via direct collapse after stellar death (as
in \citet{Alvarez2009}). The accretion rate, and thus luminosity assuming
$\epsilon_{\bullet} = 10\%$, is the Bondi--Hoyle accretion rate of the cell in
which the BH resides. The simulation volume is $0.25 \ \mathrm{Mpc} \ h^{-1}$
on a side, with $128^3$ particles and cells on the root grid. A single nested
grid occupies the inner $1/8$ of the volume at twice the root grid resolution,
where eight additional levels of adaptive-mesh refinement are allowed,
yielding a peak spatial resolution of $0.23 \mathrm{pc} \ h^{-1}$.

We run two simulations, each identical to the other except for the choice of
discrete SED. Our `control' simulation uses monochromatic SEDs --- the PopIII
star is a monochromatic source of $E = 29.6$ eV photons, while the X-ray
source emits at $E = 2$ keV. The second simulation employs the optimal
four-bin SEDs found in Tables \ref{tab:OptimalBBspectra} and
\ref{tab:OptimalPLspectra}.

\begin{figure}[ht]
\centering
\subfigure[]{
\includegraphics[width=0.48\textwidth]{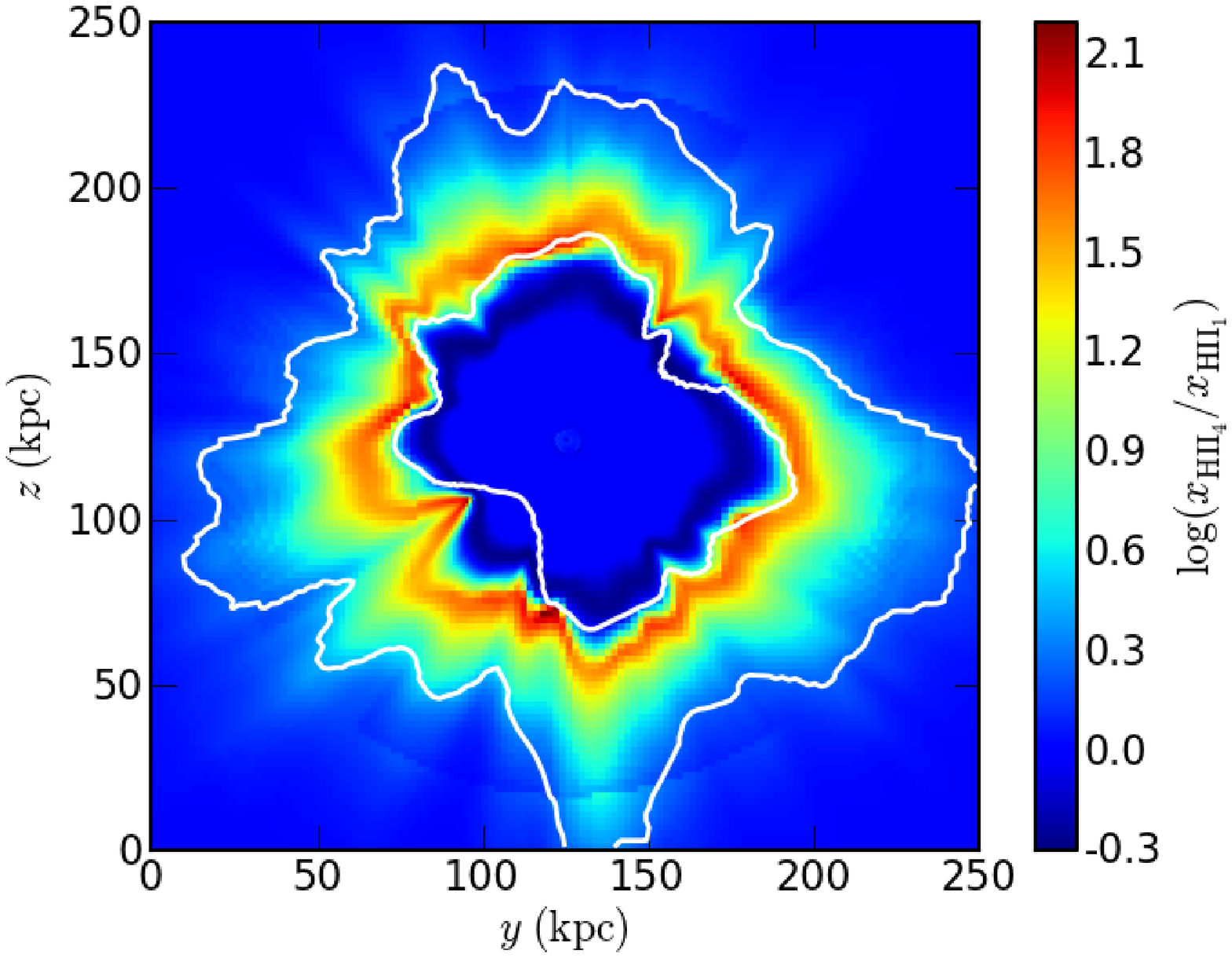}
\label{fig:xratio}
}
\subfigure[]{
\includegraphics[width=0.48\textwidth]{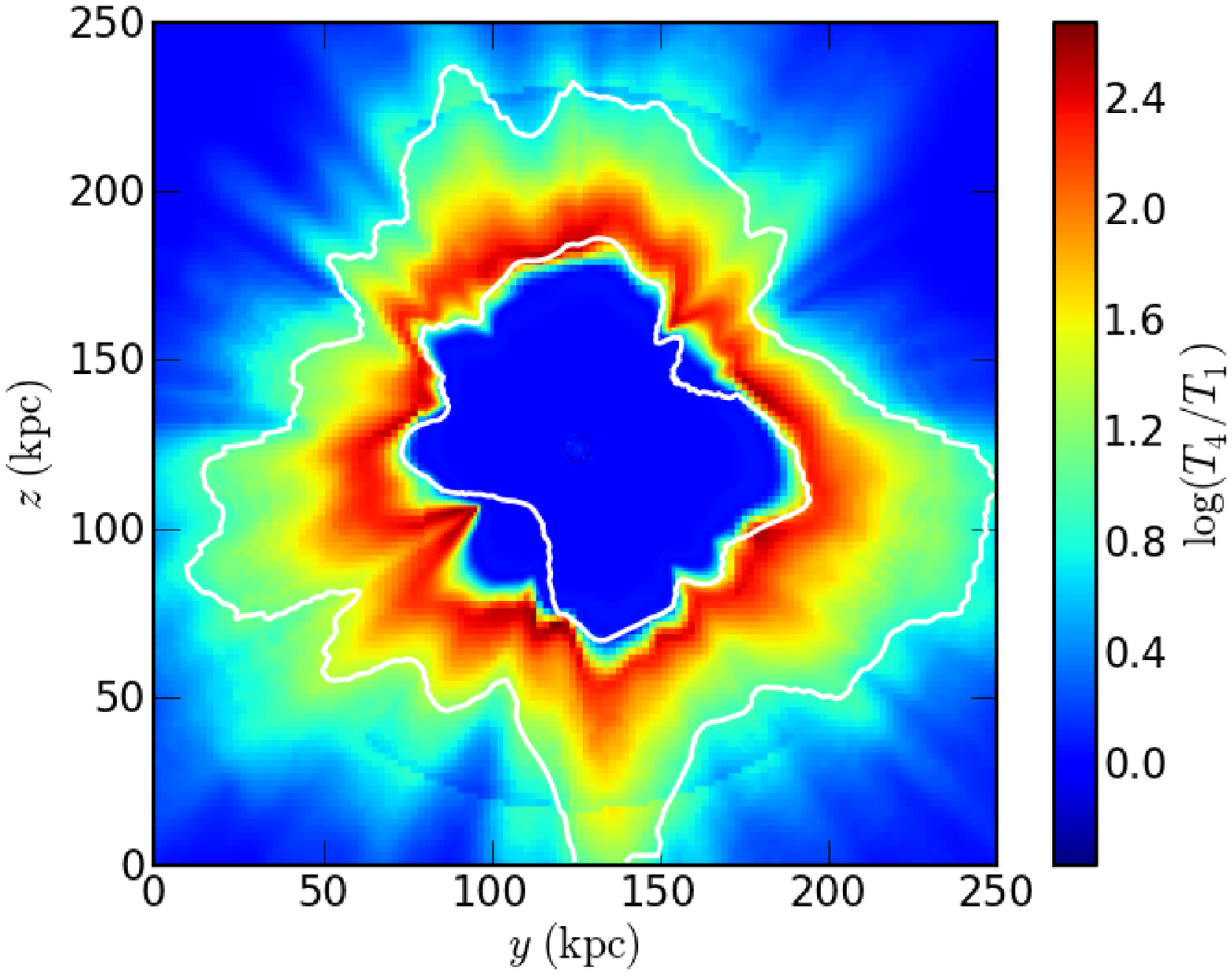}
\label{fig:Tratio}
}
\caption[]{Ratio of slices of the ionized fraction (a) and temperature (b) obtained using our optimized $n_{\nu} = 4$ blackbody SED ($x_{\text{H } \textsc{ii}_4}, T_4$) and the standard monochromatic SED ($x_{\text{H } \textsc{ii}_1}, T_1$).  Both slices are $2.25$ Myr after the formation of a Population III star.  Contours (from center outwards) correspond to hydrogen column densities of $\NHI = 2$ and $4 \times10^{19} \ \mathrm{cm^{-2}}$.}
\label{fig:EnzoResults}
\end{figure}

As shown in Figure \ref{fig:EnzoResults}, the magnitude of the errors between
monochromatic and $n_{\nu} = 4$ solutions is even more significant in the
cosmological problem than in the RT06 test problem, since the ionizing
luminosity of the blackbody source considered is nearly two orders of
magnitude larger ($1.2 \times 10^{50}$ versus $5 \times 10^{48} \
\mathrm{s^{-1}}$).  For very luminous sources, even small errors in $\Phi$ and
$\Psi$ will become noticeable as characteristic timescales for
photoionization and heating are short. 

During the BH phase of evolution, there are more ways for the monochromatic
and multi-frequency solutions to differ aside from the SEDs being employed.
The accretion luminosity depends on local gas properties, which will be
different in each simulation due to errors accrued during the PopIII star's
lifetime. Properties of the broader medium will of course vary for the same
reason, leading to changes in how far soft X-rays are able to propagate before
being absorbed. Throughout the 100 Myr of evolution after the PopIII star's
death, the Bondi--Hoyle accretion rate and thus luminosity of the accreting BH
is on average an order of magnitude smaller in the $n{\nu} = 4$ simulation
than for the monochromatic case. Errors in ionization and temperature
exceeding an order of magnitude persist throughout the BH phase as well.
Rather than attempt to disentangle the BH phase induced errors from the
preexisting errors, we simply emphasize that SED-induced errors will compound
in feedback situations like this, since the initial conditions of each
subsequent generation of objects will have been contaminated by errors
associated with the previous one.

We cannot comment on the relative errors between monochromatic and
multi-frequency treatments beyond the outermost column density contour, as our
optimization extended only to $\NHI = 3.1 \times 10^{19} \ \mathrm{cm^{-2}}$.
Future work focused on larger cosmological volumes, more luminous sources, and
harder radiation fields will need to construct optimal SEDs valid beyond $\NHI
= 10^{20} \ \mathrm{cm^{-2}}$, at least.

\section{DISCUSSION} \label{sec:Discussion}
Algorithms developed for the purpose of studying point-source radiation (e.g.,
ray-tracing) are in principle capable of propagating continuous radiation
fields, that is, tabulating Equations (\ref{eq:PHI}) and (\ref{eq:PSI}) and
computing ionization and heating rates via Equations
(\ref{eq:Gamma_PhiPsi})--(\ref{eq:Heat_PhiPsi}). The reason many have not
taken this approach could be due to the additional computational overhead
involved with using continuous SEDs --- the quantities $\Phi_i$ and $\Psi_i$
must be tabulated over the complete column density interval of interest. This
includes column densities of all absorbing species, each of which must extend
from the smallest expected column (i.e., the column density of a ``fully
ionized'' cell --- we adopted a minimum species fraction of $x_{\mathrm{min}}
= 10^{-5}$) up to the largest expected column (i.e., the column density of a
fully neutral medium). The dimensionality of $\Phi_i$ and $\Psi_i$ can be
increased even further if for example energy-dependent secondary electron
treatments \citep[e.g.,][]{Ricotti2002, Furlanetto2010} or time-dependent SEDs
are of interest.

For the simulations of Section \ref{sec:PL}, we generated three-dimensional
lookup tables for $\Phi_i$ and $\Psi_i$ covering the column density range
$10^{11} < \NHI < 10^{21}$, and $10^{10} < \NHeI,\NHeII < 10^{20}$, sampling
$\NHI$ at 200 points, and $\NHeI$ and $\NHeII$ with 100 points each, resulting
in six three-dimensional tables, each consisting of $2\times10^6$ elements. We
found that poorer sampling (e.g., tables of dimension 100 $\times$ 50 $\times$
50) leads to artificial ``notches'' in ionization and temperature profiles due
to errors in the trilinear interpolation. In our case, $\PhiHI = \PhiHeI =
\PhiHeII$ and $\PsiHI = \PsiHeI = \PsiHeII$ since all emission occurs above
$10^2$ eV, making the lower limit of integration for each quantity identical.
In the general case, where emission extends all the way to the hydrogen
ionization threshold, all six quantities would be unique. Generating these
tables can take hundreds of CPU hours or more for a single SED depending on
the number of column density elements. In addition, the radiative transfer
solver requires additional modules to read in the lookup table, and perform
interpolation four times per absorbing species per grid element (see Eqs
(\ref{eq:Gamma_PhiPsi})-(\ref{eq:Heat_PhiPsi})). For sources with discrete
SEDs, one can simply compute the photo-ionization rate for each neutral
species, from which point the secondary ionization and heating rate
coefficients are obtained in a simple algebraic fashion (see Eqs
(\ref{eq:Gamma_simple})-(\ref{eq:Heat_simple})).

For high-resolution simulations focused on a single source of radiation
\citep[e.g.,][]{Kuhlen2005,Alvarez2009}, the additional effort required to
accommodate continuous radiation fields seems well worth it to ensure that the
ionization and thermal state of the gas is captured accurately. However, in
large-scale simulations of cosmic reionization, which may spawn hundreds of
thousands or perhaps millions of radiating `star particles' (depending on the
simulation volume, resolution, etc.), ray-tracing methods are certainly not
the most computationally advantageous algorithm. This is because the
computational cost of a ray-tracing calculation scales with the number of
radiation sources \textit{and} the number of frequency bins in each source SED
\citep[though the former cost can be mitigated by merging nearby radiation
sources;][]{Trac2007,Okamoto2012}. If photons with long mean free paths are of
interest, the simulation will be even more expensive since rays must be
followed to larger distances, i.e., more ray segments and iterations of the
numerical solver are required. An appealing option is to instead use
moment-based methods such as the Variable Eddington Tensor approach
\citep[e.g.,][]{Gnedin2001,Petkova2009}, flux-limited diffusion
\citep[e.g.,][]{Reynolds2009}, or other variations
\citep{Gonzalez2007,Aubert2008,Finlator2009}, as the computational cost of
such algorithms is independent of the number of radiation sources and the mean
free paths of photons, scaling only with the number of frequency bins in each
source spectrum.

As discussed in Section \ref{sec:Introduction}, multi-group schemes common in
the literature are an improvement over fiducial discrete SEDs, though it is
not generally clear how many bandpasses are required for a given problem, or
where they should lie in frequency space. Moreover, multi-group radiation
suffers from the same problem as discrete polychromatic emission: photons at
each frequency are absorbed near a characteristic column density,
$N_{\mathrm{char}}$. Computing new spectrum-weighted absorption
cross-sections, $\bar{\sigma}_n$, for each frequency group merely shifts the
location of $N_{\mathrm{char}}$.

In principle, our minimization technique could be used to optimally select
which bandpasses should be used for a multi-group algorithm, though in
practice it would be much more computationally expensive. Rather than varying
the location ($\nu_n$) or normalization ($I_n$) of frequency bin $n$ on each
Monte Carlo step, one would instead vary the position of bandpass edges, which
would change the mean photon energy in each bandpass ($h\bar{\nu}_n$) and
spectrum-weighted cross section, $\bar{\sigma}_n$ \citep[e.g.,][]{Aubert2008}.
Because $h\bar{\nu}_n$ and $\bar{\sigma}_n$ are integral quantities, they
would need to be computed numerically on each Monte-Carlo step, and thus
hundreds of thousands of times for a single optimization.

\section{CONCLUSIONS} \label{sec:Conclusions}  
We have shown that the manner in which a discrete SED is constructed can
induce substantial errors in simulation results, both in the ionization and
temperature profiles around stars and quasars. But, these errors can be
avoided to a large degree using only four discrete emission frequencies if
source SEDs are designed via the methods of Section \ref{sec:Methods}.
Discrete SEDs constructed in a simple way (e.g., bins linearly spaced in
frequency) will perform more poorly than optimally selected SEDs with the same
number of bins, since it is the column density interval of interest that
dictates the range of photon energies required, and the power to which each is
assigned.

In general, discrete SED treatments fail to ionize and/or heat gas at large
column densities, i.e., large physical scales or environments with dense
clumps of gas. This has strong implications for simulations dedicated to
understanding the magnitude and mode of radiative feedback on gas surrounding
radiation sources. Current questions of this sort include whether or not
radiation stimulates or suppresses further star formation in nearby
proto-stellar clouds, and if radiative feedback can stifle the growth of SMBHs
at high redshift.
  
As expected, extending our one-dimensional work to three-dimensions produces
ionized regions around a first star and remnant BH that deviate significantly
in ionized fraction, temperature, size, and morphology. Such findings have
implications in radiative feedback, but also in studies of both hydrogen and
helium reionization. Certainly miscalculations of the ionization state of gas
surrounding galaxies in the early universe will lead to errors in the volume
averaged neutral fraction, volume filling factor of ionized gas, and the
optical depth of the CMB to electron scattering ($\tau_e$). As we demonstrated
in Section \ref{sec:Consequences}, such errors also introduce uncertainties in
the interpretation of future 21 cm measurements, since the primary observable
quantity ($\delta T_b$) depends directly on the hydrogen neutral fraction,
electron density, and gas kinetic temperature.

Our optimizations in this work are by no means comprehensive, having selected
two commonly used radiation sources (UV blackbody and X-ray power law) as test
cases to demonstrate the method. However, optimization for more complex
spectra is straightforward, and any new optimizations run will be made
publicly available by the authors. The minimization code and one-dimensional
radiative transfer codes are both available upon request. We leave more
detailed investigations of reionization and radiative feedback, including
multiple radiation sources and multi-frequency radiation transport, to future
work.\\

The authors thank Steven Furlanetto and Daniel Reynolds for feedback on
earlier versions of this draft, as well as the anonymous referee for a
thorough review and many helpful suggestions. The LUNAR consortium
(http://lunar.colorado.edu), headquartered at the University of Colorado, is
funded by the NASA Lunar Science Institute (via Cooperative Agreement
NNA09DB30A) to investigate concepts for astrophysical observatories on the
Moon. This work used the \texttt{JANUS} supercomputer, which is supported by
the National Science Foundation (award number CNS-0821794) and the University
of Colorado Boulder. The \texttt{JANUS} supercomputer is a joint effort of the
University of Colorado Boulder, the University of Colorado Denver, and the
National Center for Atmospheric Research.

\bibliography{references}
\bibliographystyle{apj}

\newpage
\appendix

\section{Optimization via Simulated Annealing}
To solve Equation (\ref{eq:logMinimize}), we employ the Monte Carlo method of
Simulated Annealing \citep{Kirkpatrick1983,Cerny1985}. For a given source and
\nnu, we run $K$ Monte-Carlo trials, each consisting of $L$ steps, aimed at
determining the optimal values of $I_n$ and $\nu_n$ for \nnu\ frequency bins.
We do not require the bolometric luminosity of sources to be conserved (i.e.,
$\sum_{n = 1}^{n_{\nu}} I_n \neq 1$ is allowed), since some photons may
traverse the entire one-dimensional ``volume'' without ionizing a single atom,
or some fraction of the luminosity may be emitted below the hydrogen
ionization threshold. Inclusion of such photons would be computational effort
wasted in a fully three-dimensional ray-tracing calculation, for example,
since their mean free paths are very long, and once absorbed they may
contribute negligibly to ionization and heating.

Each random walk begins with randomly generated values of $\nu_n$ distributed
between the hydrogen ionization threshold and the maximum emission frequency
in the spectrum, and randomly generated values of $I_n$ that sum to unity.
Subsequent steps vary the energy or normalization of (randomly chosen)
frequency bin $n$. In order to steer each random walk towards the global
minimum, we first evaluate the quantity \begin{equation} P =
\mathrm{exp}\left[-(f_{k,l} - f_{k,l-1}) / T_{\mathrm{SA}} \right]
\label{eq:Probability} \end{equation} where $k=0,1,2,\ldots,K$ represents the
current step in the current random walk, $l$, where $l = 0,1,2,\ldots,L$, and
$f$ is the ``cost function,'' a measure of how good our current solution is.
We adopt a cost function which is the sum of errors in $\Phi_i$ and $\Psi_i$
over the column density range of interest. For each species ($i$), and each
integral quantity ($\Phi$, $\Psi$), we add the maximum deviation from
continuous and discrete solutions in the optically thin limit (first term in
Equation (\ref{eq:Cost})), the maximum deviation over the entire column
density range (second term in Equation (\ref{eq:Cost})), and the average
deviation over the entire column density range (final term in Equation
(\ref{eq:Cost})), all in dex, i.e.,
\begin{equation}
    f_{k,l} = \sum_i \sum_{\Lambda = \Phi,\Psi} \left\{ \mathrm{max} \left[ \mathrm{log} \left(\frac{\Lambda_i}{\Lambda_i^{\prime} (\nu_{k,l}, I_{k,l})} \right)_{\tau = 0} \right] \right.  \left. + \mathrm{max} \left[ \mathrm{log} \left(\frac{\Lambda_i}{\Lambda_i^{\prime} (\nu_{k,l}, I_{k,l})} \right)_{\tau > 0} \right] \right.  + \left. \left\langle \mathrm{log} \left( \frac{\Lambda_i} {\Lambda_i^{\prime} (\nu_{k,l}, I_{k,l})} \right)_{\tau > 0}\right\rangle  \right\} . \label{eq:Cost}
\end{equation} 
At each step in a given random walk, we also generate a random number, $q
\in [0,1]$, that will determine whether we keep our current guess,
$(\nu_{k,l}, I_{k,l})$, or revert to our previous guess, $(\nu_{k,l-1},
I_{k,l-1})$. The condition for keeping our current guess is $P \ge q$.

The key aspect of this analysis is how we vary the control parameter
$T_{\mathrm{SA}}$, which is called the temperature in analogy with Boltzmann's
equation (we add the subscript SA to distinguish the gas kinetic temperature
from this unphysical Simulated Annealing temperature). Equation
(\ref{eq:Probability}) tells us that regardless of the value of
$T_{\mathrm{SA}}$, if $f_{k, l} < f_{k, l-1}$ (i.e., our most recent guess is
better than the last), then $P \ge 1$, and we have a 100\% chance of keeping
our current guess. In other words, our method of controlling the
$T_{\mathrm{SA}}$ only effects how we deal with bad guesses --- decreasing the
temperature means we become less tolerant of bad guesses. There are many ways
of doing this \citep{Press1992}, but for simplicity we adopt the following
technique. Every $s / n_{\nu}$ steps per frequency bin, we take
\begin{equation}
    T \rightarrow \lambda T ,
\end{equation}
where $\lambda$ is an experimentally determined quantity of order unity. For
all results presented here, we have adopted $\lambda = 0.98$, and $s / n_{\nu}
= 10$. We change the number of steps per random walk depending on the
dimensionality, $2n_{\nu}$. We have found through experimentation that a good
rule of thumb is $L = 5000$ steps per trial, $K$, per frequency bin \nnu\ for
our choice of $\lambda$ and $s/n_{\nu}$. These control parameters are fairly
conservative --- further experimentation with them may yield converged
solutions for fewer trials, $K$, and steps, $L$.

\section{Code Verification}
Our one-dimensional radiative transfer code solves Equations
(\ref{eq:HIIRateEquation})--(\ref{eq:TemperatureEvolution}) using the implicit
Euler method for integration and a Newton--Raphson technique for root finding.
Each simulation is initialized on a grid of $N_c$ cells between $L_0$ and
$\Lbox$, such that the finest resolution element is $\dx = (\Lbox - L_0) /
N_c$, or simply $\dx = 1 / N_c$ in code units. Gas inside of the start radius,
$L_0$, contributes no optical depth, and Equations
(\ref{eq:HIIRateEquation})--(\ref{eq:TemperatureEvolution}) are not solved. For the
purposes of this section, we chose to use $N_c$ linearly spaced cells between
$L_0$ and $\Lbox$, though our code allows arbitrarily structured grids.

In order to track the propagation of ionization fronts accurately, we limit
the time-step based on a maximum neutral fraction change as introduced in \citet{Shapiro2004},
\begin{equation}
    \Delta t_i = \epsilon_{\mathrm{ion}} \frac{n_i}{|dn_i/dt|} , \label{eq:HIIRestrictedTimestep}
\end{equation}
where we include all absorbing species, $i = $\HI, \HeI, \HeII, and set
$\Delta t = \mathrm{min}(\Delta t_i)$. We additionally require that the
time step increase by a factor of two at most, as in \citet{Wise2011}. For all
simulations presented in this work, we have set $\epsilon_{\mathrm{ion}} =
0.05$.

The primary solver implemented in our code assumes the speed-of-light is
infinite. Such an algorithm is appealing for two main reasons, aside from the
fact that it is a very good approximation for the problems presented in this
work. First, treating the speed-of-light explicitly introduces additional
computational overhead as ``photon packages'' must be launched from the
radiation source at each time step and tracked until they exit the domain. In
the earliest stages of I-front propagation, the time step can be very small
(as required by Equation (\ref{eq:HIIRestrictedTimestep})), meaning the total
number of photon packages, $N_p$, will be much larger than the total number of
grid cells, $N_c$. Whereas $c = \infty$ treatments only require Equations
(\ref{eq:HIIRateEquation})-(\ref{eq:TemperatureEvolution}) to be solved once
per cell, finite speed-of-light treatments require this system of equations to
be solved for each photon package. At later times, when $N_p < N_c$, solving
the ion and heat equations is cheaper for finite speed-of-light treatments,
though this offers no real advantage since the majority of the computational
expense is at early times when I-front propagation is fastest. We have also
included a finite $c$ solver to accommodate a broader class of problems that
may be of interest in future work.

The second advantage of assuming $c = \infty$ is that it allows the code to be
efficiently parallelized. If $c = \infty$, cells in the domain can be solved
in arbitrary order by a single processor, or simultaneously by a network of
processors, since the radiation incident on any cell is predetermined at the
outset of each individual time step. Previous authors have ensured causality
by solving cell $k$ before cell $k+1$ at time $t$ (where increasing $k$
corresponds to increasing $r$), but this is not in fact necessary ---
causality is ensured by the monotonicity of column density with distance. In
other words, when $c = \infty$, $\ncol$ does not change within any given time
step, and so the column density (and thus radiative flux) to cell $k$ is less
than the column density (and flux) to cell $k + 1$, meaning the solution of
Equations (\ref{eq:HIIRateEquation})-(\ref{eq:TemperatureEvolution}) in cell
$k + 1$ is completely independent of the properties of cell $k$ at time $t +
\Delta t$.

\begin{figure}[ht]
\centering
\subfigure[]{
\includegraphics[width=0.48\textwidth]{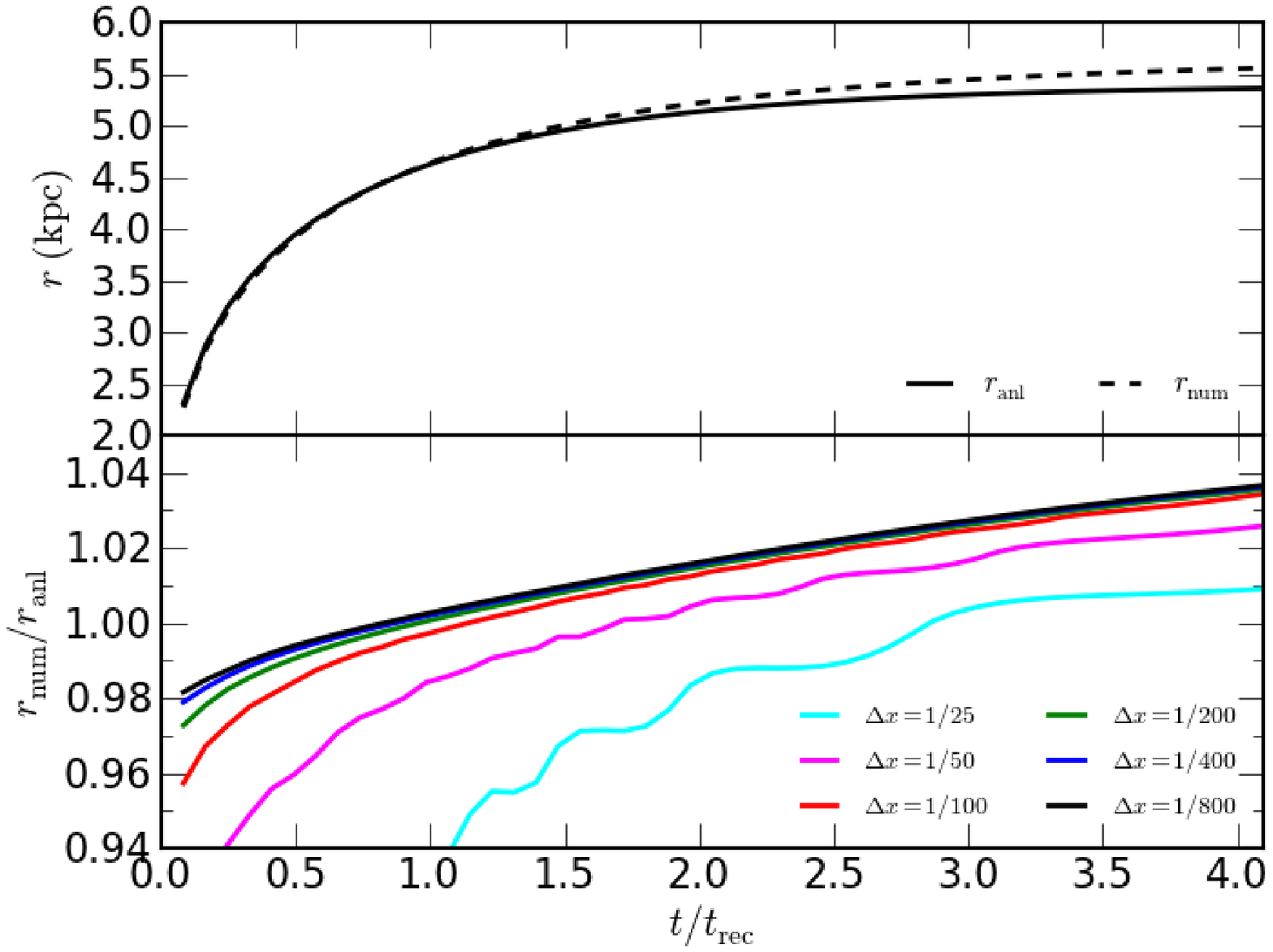}
\label{fig:RT_Test1_IfrontEvolution}
}
\subfigure[]{
\includegraphics[width=0.48\textwidth]{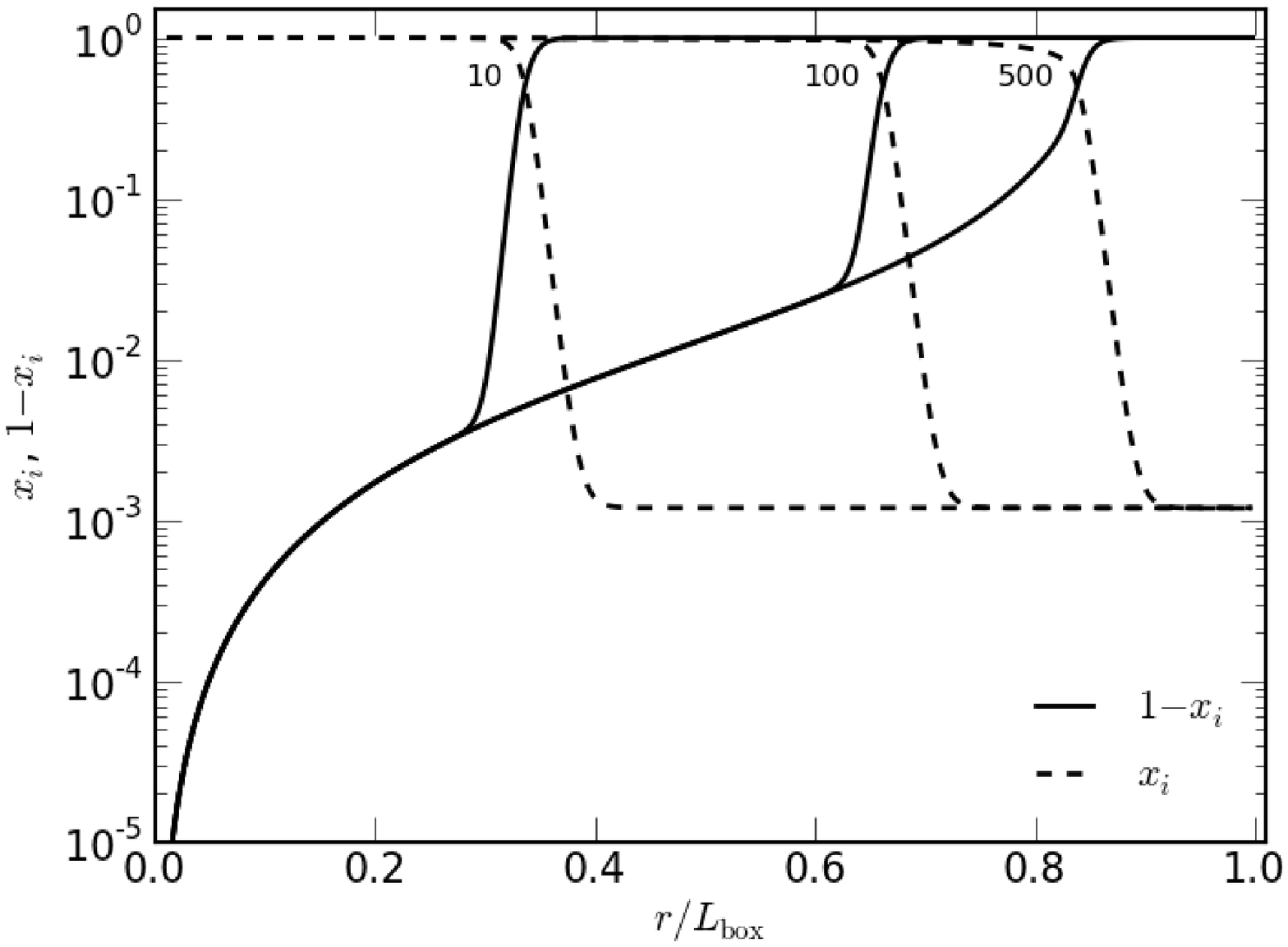}
\label{fig:RT_Test1_RadialProfiles}
}
\caption[]{Test 1: (a) Comparison of the numerical (dashed) and analytic (solid) solutions for the position of an expanding ionization front as a function of time in a hydrogen-only, isothermal medium (RT06 problem 1; top), and the ratio of the calculated and analytic solutions as a function of time and grid resolution (bottom).  The numerical solution displayed in the top panel is from the highest resolution simulation (800 grid cells, i.e., $\Delta x = L_{\mathrm{box}}/800$). ((b)) Radial profiles of the neutral (solid) and ionized (dashed) fractions at $t = 10$, $100$, and $500$ Myr.}
\label{fig:stuff}
\end{figure}

To demonstrate the functionality of the code, we repeat tests 1 and 2 from the
Radiative Transfer Comparison Project (\citet[][hereafter referred to as
RT06]{Iliev2006}) on a grid of 200 linearly spaced cells. Test 1 is the
expansion of an \HII\ region in a hydrogen-only, isothermal medium surrounding
a monochromatic source of 13.6 eV photons. We adopt the same parameters used
in RT06: constant temperature $T = 10^4 \ \mathrm{K}$, uniform hydrogen number
density $\nH = 10^{-3} \ \mathrm{cm^{-3}}$, ionized fraction $\xHII = 1.2 \times
10^-3$, in a box $L_{\mathrm{box}} = 6.6 \ \mathrm{kpc}$ in size, and with
photon luminosity $\dot{Q} = 5\times10^{48} \ \mathrm{s^{-1}}$. The classical
analytic solution for the radius of an ionization front is
\begin{equation}
    r_{\mathrm{IF}}(t) = r_s (1 - e^{-t/t_{\mathrm{rec}}})^{1/3} ,
\end{equation}    
where $r_s$ is the Str\"{o}mgren radius, 
\begin{equation}
    r_s = \left(\frac{3 \dot{Q}}{4\pi \recHII \nHI^2}\right)^{1/3} ,
\end{equation}
and the recombination time, \trec, is defined as
\begin{equation}
    t_{\mathrm{rec}} \equiv \frac{1}{\recHII \nHI} .
\end{equation}
This solution is approximate even in isothermal media, given that it assumes a
constant neutral hydrogen density, $\nHI$. More accurate analytic solutions
exist \citep{Osterbrock2006}, and predict a departure from the classical
solution at $t/t_{\mathrm{rec}} \simeq 1$, which grows to a $\sim 5 \%$
difference by $t/t_{\mathrm{rec}} \simeq 4$. Our numerical solution (see
Figure \ref{fig:RT_Test1_IfrontEvolution}) captures this behavior very well.
In Figure \ref{fig:RT_Test1_RadialProfiles}, we show radial profiles of the
ionized and neutral fractions at three stages of the I-front expansion, which
are again in very good agreement with the calculations presented in
RT06.

\begin{figure}[ht]
\centering
\subfigure[]{
\includegraphics[width=0.48\textwidth]{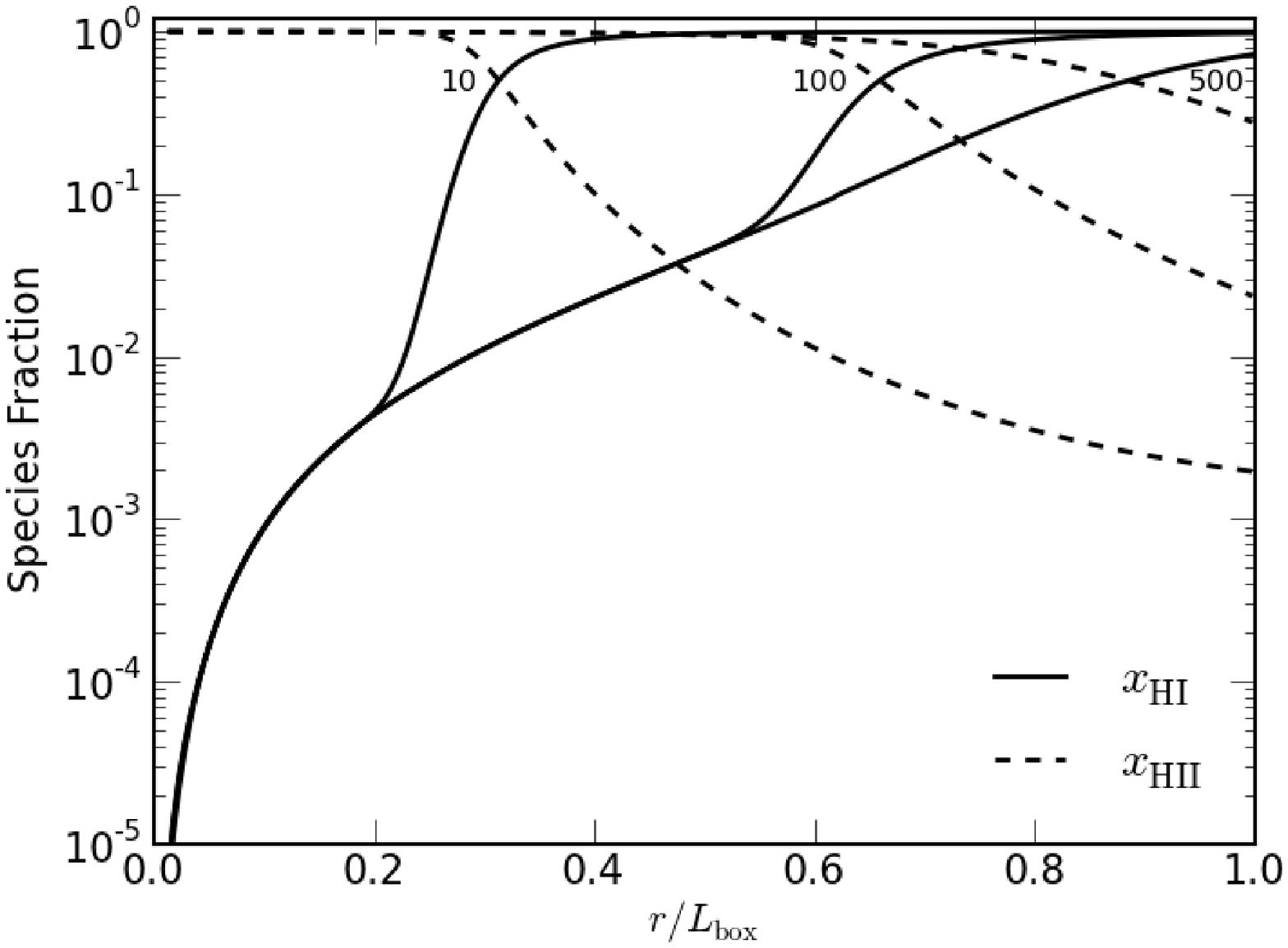}
}
\subfigure[]{
\includegraphics[width=0.48\textwidth]{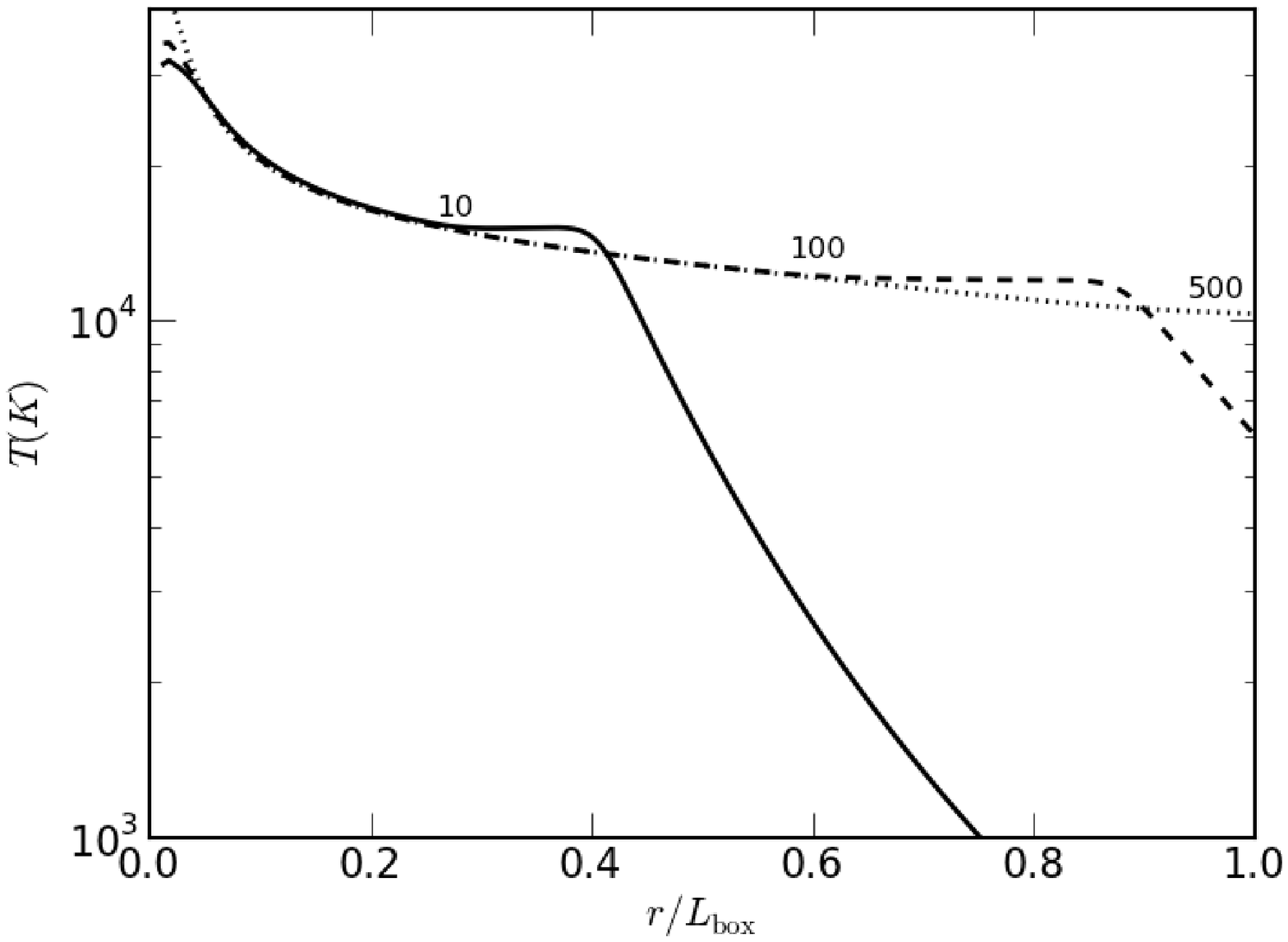}
}
\caption[]{Test 2: (a) Radial profiles of the neutral (solid) and ionized (dashed) fractions at $t = 10$, $100$, and $500$ Myr. (b) Radial profiles of the kinetic temperature at $t = 10$, $100$, and $500$ Myr (solid, dashed, and dotted lines, respectively).}
\label{fig:RT_Test2_RadialProfiles}
\end{figure}

Test 2 is the same as Test 1, except now the temperature is allowed to evolve
according to Equation (\ref{eq:TemperatureEvolution}), and the monochromatic
radiation source is replaced by a $10^5$ K blackbody spectrum. Radial profiles
of the neutral and ionized fractions and temperature can be seen in Figure
\ref{fig:RT_Test2_RadialProfiles}. Again, our numerical solutions are in very
good agreement with previous work.  

\end{document}